\documentstyle[floats,aps,prb,epsfig,eqsecnum,amsmath,amssymb]{revtex}

\renewcommand{\d}{{\mbox{d}}}

\newcommand{\bs}{\baselineskip}

\newcommand{\bi}{\begin{itemize}}
\newcommand{\ei}{\end{itemize}}
\newcommand{\bq}{{\bf q}}

\newcommand{\bu}{{\bf u}}

\newcommand{\bG}{{\bf G}}

\newcommand{\bk}{{\bf k}}
\newcommand{\bz}{{\bf z}}
\newcommand{\bN}{{\bf 0}}
\newcommand{\cH}{{\mathcal{H}}}
\newcommand{\cG}{{\mathcal{G}}}

\newcommand{\cO}{{\mathcal{O}}}
\newcommand{\bP}{{\sf P}}
\newcommand{\bR}{{\bf R}}
\newcommand{\bX}{{\bf X}}
\newcommand{\bY}{{\bf Y}}
\newcommand{\bx}{{\bf x}}
\newcommand{\br}{{\bf r}}

\newcommand{\bD}{{\sf D}}
\newcommand{\bM}{{\sf M}}

\newcommand{\nn}{\nonumber}
\newcommand{\btu}{\tilde{{\bf u}}}
\newcommand{\btx}{\tilde{{\bf x}}}
\newif\ifbig
\bigfalse

\ifbig
        
\fi

\begin{document}
\draft
\date{\today}

\ifbig
{}
\else
\twocolumn[\hsize\textwidth\columnwidth\hsize\csname@twocolumnfalse%
\endcsname
\fi
%------------------------- title  ----------------------------------
\title{Nonuniversal Correlations and Crossover Effects in the Bragg-Glass Phase of
Impure Superconductors}
\author{Simon Bogner${}^1$
\and Thorsten Emig${}^2$
\and Thomas Nattermann${}^1$}
\address{${}^1$ Institut f\"ur Theoretische Physik, Universit\"at zu
        K\"oln, Z\"ulpicher Stra\ss e 77, D-50937 K\"oln, Germany 
        \\
        ${}^2$Physics Department, Massachusetts Institute of Technology,
        Cambridge, Massachusetts 02139, USA 
}

\maketitle
\begin{abstract}
  The structural correlation functions of a weakly disordered
  Abrikosov lattice are calculated in a functional RG-expansion in
  $d=4-\epsilon$ dimensions.  It is shown, that in the asymptotic
  limit the Abrikosov lattice exhibits still quasi-long-range
  translational order described by a {\it nonuniversal} exponent
  $\eta_{\bf G}$ which depends on the ratio of the renormalized
  elastic constants $\kappa ={c}_{66}/ {c}_{11}$ of the flux line (FL)
  lattice. Our calculations clearly demonstrate three distinct scaling
  regimes corresponding to the Larkin, the random manifold and the
  asymptotic Bragg-glass regime.  On a wide range of {\it
    intermediate} length scales the FL displacement correlation
  function increases as a power law with twice the manifold roughness
  exponent $\zeta_{\rm RM}(\kappa) $, which is also {\it
    nonuniversal}.  Correlation functions in the asymptotic regime
  are calculated in their full anisotropic dependencies and 
  various order parameters are examined.  Our results, in particular the
  $\kappa$-dependency of the exponents, are in variance with those of
  the variational treatment with replica symmetry breaking which
  allows in principle an experimental discrimination between the two
  approaches.
\end{abstract}

\ifbig
{}
\else
]
\fi

%-------------------------- text ---------------------------------
%
%---------------------------
%---------------------------
\section{Introduction}
\label{intro}
%---------------------------

%---------------------------

The discovery of high-$T_c$ superconductors in 1986 by Bednorz and
M\"uller \cite{Muller86} has led to a strongly renewed interest in the
theory of superconductivity, both in the explanation of its
microscopic origin \cite{Anderson} and in the application of the
phenomenological theory on the determination of the phase diagram,
of dissipation effects due to flux creep and related phenomena
\cite{Blatter+94,Nattermann+00}.

Thermal fluctuations turned out to lead to a phase diagram drastically
different from the mean-field prediction even in pure systems.  This
is due to the elevated transition temperatures as well as due to the
pronounced layer structure of the high-$T_c$ materials.  A liquid
phase is now expected to separate the flux repulsing Meissner from the
Abrikosov phase. In the latter the magnetic induction ${\bf B}$ enters the
material in the form of quantized flux lines (FLs) which still form a
triangular lattice. However, the upper critical field $H_{c_2}$ (where
superconductivity disappears) is now strongly reduced with respect to its
mean-field value and can be understood as resulting from a melting of
the flux line lattice \cite{Blatter+94}.  
In fact there are several experiments on
high--$T_c$ materials providing firm evidence that the Abrikosov
lattice is melted over a significant region of the phase diagram
\cite{Cubitt+93,Zeldov+95}.  The field dependent melting line can be
obtained from Lindeman's criterion in conjunction with a detailed
knowledge of the elastic properties of the lattice.  In general, these
properties have to be obtained by methods going beyond a simple
continuum description \cite{Miguel+00}.  Moreover, at present, it is
not clear, whether the transition to the normal phase at high fields
happens in these materials via one or two transitions.

It is well known that in addition to thermal fluctuations in type--II
superconductors also the effect of disorder has to be taken into
account since FLs have to be pinned in order to prevent dissipation
from their motion under the influence of an external current.
Therefore, understanding the interaction of vortices with quenched
randomness in form of these defects is of especial importance. 

A natural question arising in this context concerns the influence of
randomness on the translational order of the lattice.  This question
was first considered by Larkin\cite{Larkin70} who found from
perturbation theory that randomly distributed pinning centers lead
indeed to a destruction of the Abrikosov lattice. In
particular, he obtained an {\it exponential} decay of the correlations of
the order parameter for translational long range order $\Psi_{\bf
  G}({\bf r}) = e^{i{\bf Gu}({\bf r})}$ on length scales larger than a
disorder dependent Larkin length $L_{\xi }$. Here ${\bf G}$ and ${\bf
  u}$ denote a reciprocal lattice vector and the displacement field of
the FL lattice, respectively, and $\br=(\bx,z)$ is a 3-dimensional
position vector.

This conclusion was in agreement with the more general observation of
Imry and Ma, that quenched randomness coupled to a continuous symmetry
order parameter destroys true long range order in less than four
dimensions \cite{Imry+75}.  In this context the Imry-Ma argument played 
a similar role as the Mermin-Wagner theorem for pure system in two
dimensions. In the latter case we know however, that the destruction of
true long range order by thermal fluctuation contains still the
possibility of topological order with an algebraic decay of
correlations \cite{KosterlitzThouless73}.
The intriguing possibility of a weak disorder phase with a
topological order, which distinguishes it from the fully disordered
phase at larger disorder strength, remained as an open question, 
and was discussed intensively in recent works.

As was first shown by Nattermann
\cite{Nattermann90}, in treating the interaction between FL lattice
and disorder, it is crucial to keep the periodicity of this
interaction under the transformation ${\bf u} \rightarrow {\bf u}+{\bf
X}$, where ${\bf X}$ is a lattice vector of the Abrikosov
lattice. This symmetry, which is abandoned in perturbation theory
\cite{Larkin70} and in the so-called manifold models
\cite{Bouchaud+92c}, leads to a much slower, {\it logarithmic}
increase of the elastic distortions, measured by the displacement
correlation $B(\br)=\langle[\bu(\br)-\bu(\bN)]^2\rangle$
\cite{Nattermann90,Korshunov93,Giamarchi+94,Giamarchi+95}.
If large dislocation loops can be neglected -- as assumed in the above
discussion -- then the resulting phase is characterized by a structure
factor with power-law singularities corresponding to a
quasi-long-range ordered flux phase, the "Bragg-glass"
\cite{Giamarchi+94,Giamarchi+95}, with algebraic decaying pair
correlation function $C_{\bf G} ({\bf r})=\langle\Psi_{\bf G}({\bf
r})\Psi_{-\bf G}({\bf 0})\rangle$.  

In three dimensions there is indeed
strong theoretical
\cite{Kierfeld+97,Gingras+96,Carpentier+96,Ertas+96,Kierfeld98,FisherDS97}
and experimental \cite{Kim+99} support of a phase transition from a
fully disordered phase, where the occurrence of unbounded dislocation
loops leads to an instability of the Bragg-glass, to a
dislocation-free genuine glass phase at weak disorder.

The resulting power law decay of $C_{\bf G} ({\bf r})$ in this
Bragg-glass is reminiscent of the situation in pure 2D-crystals where
in the solid phase 
$C_{\bf G} ({\bf x_{\perp}})\sim |{\bf x_{\perp}}|^{-\eta_{{\bf G}}}$.  
This solid phase corresponds in fact
to a {\it line} of critical points with 
$\eta_{{\bf G}}=TG^2(1+{\kappa}^{-1})/({4\pi}{\tilde c}_{11})$ and 
$\kappa = {\tilde c}_{66}/{\tilde c}_{11}$. The ${\tilde c}_{ii}$ represent the
{\it renormalized} elastic constants which have a finite temperature
dependent value. {\it At} the melting temperature $T_m$ the exponent
$\eta_{{\bf G}}$ reaches a non-universal value $\eta_{\bf
G}=(Ga/4\pi)^2(1-{\kappa ^2})$, which still depends on $\kappa(T_m)$
\cite{Nelson83}.

Based on this rough analogy one could expect a similar non-universal
behaviour for the Bragg-glass phase, although it is dominated by
randomness and, therefore, by a zero temperature fixed point.  But, in
addition to the relevance of metastable states, the accuracy of
earlier renormalization group approaches and variational techniques
was particularly hampered by the complex elastic properties of the
Abrikosov lattice.  For instance, Giamarchi and Le Doussal
\cite{Giamarchi+94,Giamarchi+95} calculated $C_{\bf G} ({\bf x})$ using
(i) a variational treatment for the triangular FL lattice and (ii) a
functional renormalization group (FRG) in $d=4-\epsilon$ dimensions
for a simplified model using a scalar displacement field $u$ with
isotropic elasticity only.  In both cases they found
$C_{\bf G} ({\bf x}_{\perp},0)\sim |{\bf x_{\perp}}|^{-\eta_{\bf G} }$ 
with a {\it universal} exponent ${\eta}_{\bf G_0} =A(4-d)$ with $A=1$ and
$A=\pi^2/9\approx 1.1$ for the treatment (i) and (ii), respectively.
Here ${\bf G}_0$ denotes one of the smallest reciprocal lattice
vectors with $G_0a=4\pi/\sqrt{3}$, and $a=(2\phi_0/\sqrt{3}B)^{1/2}$
is the lattice spacing. To date, the question to which extent these
results depend on the applied techniques and the simplifications of
the considered models is not fully resolved.

In this paper, we systematically study the structural properties of
the triangular Abrikosov lattice at weak disorder using a FRG
method. Contrary to previous approa\-ches, we explicitely take into
account the triangular symmetry of the lattice and all of its elastic
modes. We derive functional recursion relations for the correlation
function of the random potential, which, in combination with a Fourier
decomposition technique, allow us to extract detailed information
about the collective wandering behaviour of the FLs.  This transversal
wandering can be characterized by the roughness exponent $\zeta$
controlling the displacement correlations via 
$B(\br)\sim |\br|^{2\zeta}$. Following the RG flow, three different scaling
regimes can be clearly identified: On length scales smaller than the
Larkin length, the FL are displaced by an amount smaller than the
characteristic scale $\xi$ of the short range correlated random
potential. Thus, Larkin's perturbation theory holds with
$\zeta=(4-d)/2$. Beyond this regime, but for line displacements still
smaller than their distance $a > \xi$, one enters an intermediate
regime with non-trivial roughness influenced by metastable
states. Although this regime is similar to a single FL system, the
complex elastic interactions of the lattice lead to a {\it
non-universal} $\zeta$ depending on the ratio $c_{66}/c_{11}$ as
reported here for the first time. Finally, on asymptotic scales, the
FL displacement becomes larger than $a$, leading to the logarithmic
roughness responsible for the quasi-order of the Bragg-glass phase.
Using our Fourier technique to analyse the FRG flow equations, we were
able to determine from the fixed point value $\Delta^*$ of the
variance of the random potential the exponent $\eta_{\bf
G}=\Delta^*(Ga)^2$ with sufficient accuracy to rule out
universality. Instead, $\eta_{\bf G}$ depends also on the ratio
$c_{66}/c_{11}$. The situation is therefore indeed qualitatively
similar to that of 2D crystals {\it at} the melting temperature as
speculated above.  With the ratio $c_{66}/c_{11}$ depending in general
on ${\bf B}$ and $T$, the observation of a field-dependent $\eta_{\bf
G}$ would yield the opportunity to judge the validity of different
approximation schemes under debate \cite{Balents+96}.  A brief
description of our combination of FRG techniques and numerical
calculations and these results appeared earlier \cite{Emig+99}.

The paper is organized as follows. First, we will introduce the model
and examine its symmetries in Section 2. The derivation of the FRG
flow equations and its numerical treatment is outlined in Section 3.
A detailed discussion of the different scaling regimes and its
correlation functions is given in Section 4. In this Section we also
study correlation functions of various order parameters to obtain
information beyond the translational order parameter as a signature of
residual order at weak disorder. Experimental implications are
presented briefly at the end of Section 4.  Finally, in the last
Section we summarize and discuss our central results. Lengthy
computation is documented in the appendix.

%---------------------------
%---------------------------
\section{The Model}
\label{model}
%---------------------------
%---------------------------

%---------------------------
\subsection{The Hamiltonian}
\label{hamiltonian}
%---------------------------
In order to examine the influence of disorder on the flux line lattice
we employ the elastic description reviewed in Ref. \onlinecite{Blatter+94}.
The degrees of freedom are the positions of the flux lines $\{{\bf
  R}_{\bf X}(z)\}$ in the $(x,y)$-plane as a function of $z$, the
direction of the external magnetic field, and $\bX$, the lattice
vectors of the perfect triangular Abrikosov lattice. To measure the
disorder induced distortion of the flux lines, we introduce the
displacements $\bu_\bX(z)=\bR_\bX(z)-\bX$ from the perfect lattice.
Such a labelling with the perfect lattice positions is possible
provided no dislocations are present in the system. The stability
against dislocations has to be assured a posteriori, which gives a
finite limit for the disorder strength
\cite{Gingras+96,Kierfeld+97,Ertas+96,Carpentier+96,FisherDS97,Kierfeld98}.

In an elastic continuum description the displacements are extended
smoothly to a continuous function $\bu_\bX(z) \rightarrow\bu(\br)$
with $\br=(\bx,z)$.  Symmetry reasons require three distinct elastic
coefficients \cite{Landau_Lifschitz7} and the energy cost of a
distortion of the triangular line lattice is
\begin{eqnarray}
\label{hamx}
\cH_{el}&=&{1 \over 2}\int \!\d^2x \d z \left\{c_{11}
\left(\nabla_{\perp}  \bu \right)^2
        +c_{66}\left(\nabla_{\perp} \times \bu \right)^2+\right.  \nn\\
&&\hspace{2.2cm}\left.       
 +c_{44}\left(\nabla_z \bu \right)^2 \right\}.
\end{eqnarray} 
The elastic coefficients correspond to compression ($c_{11}$), shear
($c_{66}$) and tilt ($c_{44}$) of the lattice.  The former two
describe flux line interaction, while the tilt modulus contains both
an interaction contribution and individual line tension.  Such an
elastic continuum description is valid as long as both displacements
and their gradients vary slowly over lattice steps, which can 
 be subsumed in the assumption of $L_a \gg a$.  Here, $L_a$ is
the positional correlations length, defined as the distance of two
FLs whose mean displacements vary by the order of one lattice
spacing $a$. In our context this is a condition on the weakness
 of disorder. Below we show that it can be easily
satisfied by realistic impurity concentrations.

In wavenumber space, the elastic energy reads
\begin{eqnarray}
\label{hamq}
\cH_{el}&=& {1 \over 2} \int\limits_{\mbox{\tiny BZ}}\! 
{{\d^2q_\perp \d^{d-2} q_z}\over{(2\pi)^d}} 
\bu(\bq)\left( \cG^{-1}_L \bP_L + \cG^{-1}_T \bP_T \right) \bu(-\bq).
\nonumber\\
\end{eqnarray}
The integration runs over the Brillouin zone (BZ) and
$\bP_L^{\alpha \beta}=\frac{q_{\alpha}q_{\beta}}{q_{\perp}^2}, \quad
\bP_T^{\alpha \beta}=1-\frac{q_{\alpha}q_{\beta}}{q_{\perp}^2}$ are
projectors onto the longitudinal and transversal modes, respectively,
with propagators
\begin{eqnarray}
\label{el-propagator}
\cG^{-1}_L&=&c_{11} q_{\perp}^2 + c_{44} q_z^2 \nn\\
\cG^{-1}_T&=&c_{66} q_{\perp}^2 + c_{44} q_z^2.
\end{eqnarray}
This Fourier space formulation is more general than the model
(\ref{hamx}) since it both allows for wavenumber dependent elastic
moduli and does not include the continuum limit as only wavevectors
within the first Brillouin zone are included. The derivation of the
elastic moduli from Ginzburg-Landau theory \cite{Brandt77a,Brandt77b}
indeed gives $q$-dependent elastic constants on scales smaller than
the London penetration length $\lambda$. However, this only leads to
weakly renormalized {\em constants} on scales larger than $\lambda$.
For the case of weak disorder considered here, we focus on the
behaviour on scales far beyond $\lambda$ and may thus content
ourselves with the local version (\ref{hamx}). The coefficients
$c_{ii}$ are understood to be the renormalized ones.  In Eq.
(\ref{hamq}) we have formulated the model in general dimensions and
extended the $z$-direction to a $d\!-\!2$ dimensional space. This will
allow for an expansion around $d\!=\!4$ in the renormalization
procedure below.

The interaction of flux lines with impurities competes with elasticity and tends to roughen
the lattice. It is modelled by the
coupling 
\begin{eqnarray}                                           
\label{Hdis}
\cH_{dis}&=&\int\! \d^d\!r\: \rho_{\mbox{\tiny FLL}}(\br) V_{dis}(\br) \nn\\
&=&
\sum_{\bX} \int\! \d z^{d-2} \!\:V_{dis}\left(\bX+\bu\left(\bX,z\right),z\right)
\end{eqnarray}
of the flux line density
\begin{equation}
\label{density}
\rho_{\mbox{\tiny FLL}}(\br)=\sum_{\bX} \delta \!\left(\bx- \bX +
        \bu\left(\bX,z\right)\right)
\end{equation}
to the disorder potential $V_{dis}(\br)$ with short range correlations
on the scale $\xi=\max(\xi_{sc}, \xi_{dis})$, where $\xi_{sc}$,
$\xi_{dis}$ are the superconductor correlation length and the
correlation length of disorder density fluctuations, respectively.
The disorder potential is taken to be a random variable with Gaussian
distribution. It is characterized by
\begin{equation}
\label{random:potential}
\overline{V_{dis}(\br)}=0\:, \quad \overline{V_{dis}(\br)\,V_{dis}(\br')}=
        \Delta_{\xi}(\bx-\bx')\,\delta(z-z'),
\end{equation}
where $\Delta_{\xi}(\bx)$ is a delta-function smeared out over a
region of size $\xi$.

The continuum limit of Eq. (\ref{Hdis}) has to be taken with some
care.  Slow variations of the displacements do not suffice to allow a
straightforward continuum limit. The potential $V_{dis}(\br)$ varies
rapidly over lattice steps since $\xi\ll a$. Therefore, in the
Hamiltonian rewritten using Poisson's summation formula
\[
\sum_{\bX}f(\bX)=\int \!\d^2\!x f(\bx)[1+\sum_{\bG\ne \bN}e^{i\bG\bx}]
\]
the terms corresponding to reciprocal lattice vectors $\bG \ne \bN$
are important for the pinning energy and have to be considered
adequately \cite{Nattermann90,Giamarchi+94}.
\begin{eqnarray}
  \label{with_poisson}
  \cH_{dis} &=&  \rho_0 \int\limits_{z} \!\!\! \int\limits_{\bx} 
  V_{dis}\left(\bx+\bu\left(\bx,z\right),z\right)
  [ 1+\sum_{\bG\ne \bN}e^{i\bG\bx}]
  \nn \\
  &=& 
  \rho_0 \int\limits_{z} \!\!\!\int\limits_{\btx} 
  \left|\frac{\partial \bx}{\partial \btx}\right| 
  V_{dis}(\btx,z)[ 1+\!\!\sum_{\bG\ne \bN}
  e^{i\bG[\btx-\btu(\btx,z)]}],
\end{eqnarray}
where $\rho_0=B/\Phi_0$ is the average FL density. In the last
equation, we have substituted
\begin{eqnarray}
\label{substitute}
\btx&\equiv&\bx+\bu(\bx,z)\nn\\
\btu(\btx,z)&=&\btu(\bx+\bu(\bx,z),z)\equiv\bu(\bx,z).
\end{eqnarray}
With $\btu(\btx)$, the displacement field is now written as a function
of the actual position $\btx$ of the flux lines rather than as a
function of the perfect reference lattice coordinates $\bx$.  Such a
relabelling is possible provided that $|\nabla u_\alpha|< 1$, which is
assured by the above assumption of $|\nabla u_\alpha| \backsimeq {a/
  L_a} \ll 1$.

The change of parameterization in Eq. (\ref{substitute}) from $\{\bx,
\bu\}$ to $\{\btx,
\btu\}$ shall also be performed in the elastic energy via
\begin{eqnarray}
  \left|\frac{\partial \bx}{\partial \btx}\right| &=& \det[\delta_{\alpha\beta} - \partial_\alpha \tilde{u}_\beta]
%  \d \bx &=& \d \bx' \det[\delta_{\alpha\beta} - \partial_\alpha u'_\beta]
 \nn \\ 
  \partial_\alpha u_\beta &=& \partial_\gamma \tilde{u}_\beta[\delta_{\alpha\gamma} + \partial_\alpha u_\gamma].
\end{eqnarray}
This gives one formally invariant term plus terms of higher order in
derivatives of $u$. The latter are -- due to the extra gradients --
irrelevant in the RG to follow and therefore neglected in the first
place. The same reasoning holds for the extra gradient terms resulting
from the change of measure in the partition function of the problem,
where originally one sums over the configurations $\bu(\bx)$.  With
this in mind, we drop from now on the tilde, keep the elastic lattice
energy unchanged and derive a compact form for the pinning energy.
Retaining in Eq. (\ref{with_poisson}) only the lowest order
contributions in $\partial_\alpha u_\beta$ and subtracting
$\bu$-independent terms, we obtain
\begin{equation}
\label{h_dis2}
\cH_{dis}= \rho_0 \int_\br \big\{ -V_{dis}(\br)\nabla_\perp \bu(\br) + 
V_{dis}(\br)\sum_{\bG\ne \bN}e^{i\bG[\bx-\bu(\br)]}\big\}.
\end{equation}
The first term couples the divergence of the displacement field to the
disorder potential. It can be shown to lose against elastic energy in
the effort to roughen the lattice above two dimensions by a simple
scaling argument. One assumes the system to be rough and the
displacement to vary with the scale as $u\sim L^\zeta, \zeta<1$. The
elastic energy then scales as $L^{d-2+2\zeta}$ and the pinning energy
from the first and second term in Eq. (\ref{h_dis2}) scale as
$L^{(d-2+2\zeta)/2}$ and $L^{d/2}$, respectively. Therefore, for $d>2$
the first term can be neglected against the elastic energy, whereas
{\em all} terms of the second term are relevant for $d<4$.  The
relevance of these infinitely many exponential operators in Eq.
(\ref{h_dis2}) makes a {\em functional renormalization group}
inevitable. The preceding discussion shows that the `corrections' to
the naive continuum limit are dominant and retaining the lattice
structure is crucial (see also Section \ref{symmetries} (c)).

Consequently, we confine our analysis to the last term in Eq.
(\ref{h_dis2}). The slowly varying displacement field in the harmonics
rather than in the disorder potential itself better shows which terms
are relevant. This is made use of when we average over the disorder
potential, which is done by the standard replica method. The averaging
process introduces an interaction between different replicas, which is
calculated in  detail in Appendix \ref{Replica_pinning_energy}.
The resulting replica Hamiltonian, which will be the starting point
for the renormalization group (RG) below, reads
\begin{eqnarray}
\label{repham-complete}
\cH^n &=& {1 \over {2 }} \sum_{a,b=1}^n \int_{\bq \in BZ} \bu^a(\bq)
\left( \cG^{-1}_L \bP_L + \cG^{-1}_T \bP_T \right) \bu^a(-\bq)\, \delta_{a b}
\nn \\
&&-{1 \over 2 T} \sum_{a,b=1}^n \int \!\d^d r \:R(\bu^a(\br) - \bu^b(\br)), 
\end{eqnarray}
with the disorder correlation function defined by
\begin{equation}
\label{correlator}
R(\bu) \equiv \rho_0^2 \sum_{\bG\ne \bN} \tilde{\Delta}(\bG) e^{-i \bG \bu}.
% =\rho_0 \sum_{\bX} \Delta(\bu - \bX) .
%- \rho_0 \tilde{\Delta}(\bN)
\end{equation}
The random potential correlator $\Delta_\xi(\bx)$ [with Fourier
transform $\tilde{\Delta}(\bk)$, see Eq.  (\ref{random:potential})] is
given in terms of physical quantities by
\begin{equation}
\label{physicalDelta}
\Delta_\xi(\bx)=f^2_{pin} n_{imp} \xi^4 g(\bx/\xi),
\end{equation}
where $f_{pin}$ is the mean individual impurity pinning force,
$n_{imp}$ the impurity density and $g(\bx/\xi)$ a function of
amplitude $1$ and range $\xi$, cf. Ref. \onlinecite{Blatter+94}.  This
allows to model the bare, unrenormalized correlator as
\begin{equation}
\label{physicalR}
R_0(\bu)=f^2_{pin} n_{imp} \xi^6 \frac{B^2}{\Phi^2_0}\sum_\bG 
\Theta(1-G\xi)e^{i\bG\bu}.
\end{equation}
The second derivatives $R_{\alpha\beta} \equiv
\partial_{u_\alpha}\partial_{u_\beta}R$ of $R(\bu)$ at the origin will
be of central interest below.  Using Eq. (\ref{physicalR}) its
unrenormalized bare value can be approximated by
$R_{xx}(\bN)=R_{yy}(\bN)\backsimeq 10^2f^2_{pin} n_{imp} \xi^2
B/\Phi_0\left(1+\cO(\xi/a)\right)$.
%---------------------------
\subsection{Symmetries}
\label{symmetries}
%---------------------------
Next we want to examine the symmetries of the system in
Eq. (\ref{repham-complete}) and their constraints on the structural
order of the flux line lattice.
\begin{itemize}
\item[(a)] $\cH_{el}$ is isotropic in the ($x,y$)-plane.
  For the special choice of $c_{11}=c_{66}$ one can make $\cH_{el}$
  isotropic even in $(\bx,z)$ by rescaling $z\to z \sqrt{c_{44}/
    c_{11}}$.  Previous RG approaches \cite{Giamarchi+94} were limited to
  this very special choice ignoring anisotropy.
\item[(b)] The elastic energy $\cH_{el}$ is invariant under
  simultaneous rotation of displacements by 90 degrees $(u_x(\br),
  u_y(\br))\rightarrow (u_y(\br), -u_x(\br))$ and $\kappa \equiv
  c_{66}/c_{11} \rightarrow \kappa^{-1}$.  Divergenceless shear
  strains are rotated into curlfree compressional strains, compensated
  for by the exchange of the respective moduli.  In wavenumber space
  the longitudinal modes are exchanged with the transversal ones.
\item[(c)] The disorder correlation function $R(\bu)$ (see Eq.
  (\ref{correlator})) has the full symmetry of the triangular lattice,
  i.e., it shows invariance under translations of $\bu$ by lattice
  vectors and rotations of $\bu$ by integer multiples of 60 degrees.
  The former is obvious from $R(\bu)$ being a Fourier sum while the
  point group symmetry of $R(\bu)$ can be seen easily if one applies a
  rotation $\bD$, that maps the lattice onto itself, and uses the
  isotropy of disorder
\begin{eqnarray*}
R(\bD\bu)&=&\rho_0^2\sum_{\bG}\tilde\Delta(\bG)e^{-i\bG\bD\bu}\\
&=&\rho_0^2\sum_{\bD\bG'}\tilde\Delta(\bD\bG')e^{-i\bG'\bu}\\
&=&\rho_0^2\sum_{\bG'}\tilde\Delta(\bG')e^{-i\bG'\bu}.
\end{eqnarray*}
While disorder is assumed to be homogenously and isotropically
distributed, the reference state of the FL positions shows the
discrete triangular symmetry.  Together, for the displacements
measured from the lattice, this gives a disorder correlator with
triangular symmetry.  Due to the discrete 6-fold rotational invariance
the function $R(\bu)$ becomes isotropic at the origin, i.e.,
$R_{xx}(\bN)=R_{yy}(\bN)$, $R_{xy}(\bN)=0$.
\item[(d)] The effective replica pinning energy is invariant under
\begin{equation}
  \label{galilei}
  \bu^a(\br)\rightarrow\bu^a(\br)+\mathbf{f}(\br)
\end{equation}
for an arbitrary (yet constant in the replica index $a$) function
$\mathbf{f}(\br)$.  This is a consequence of locality of our compact
form for $\cH_{dis}$ and thus only an approximate symmetry. As can be
seen from the detailed analysis in appendix
\ref{Replica_pinning_energy}, locality grows if $\Delta_\xi(\bx)$ gets
sharper. Upon renormalization $\Delta_\xi(\bx)$
becomes more and more delta-function like. This is not surprising as
the short scales, whose coupling by a finite correlation gives the
non-locality, are integrated out.  On larger scales we may thus take
symmetry (\ref{galilei}) as given.
\end{itemize}
These symmetries allow for important conclusions, which simplify the
calculations to follow:
\begin{itemize}
\item[(i)] Symmetry (d) grants that the elastic constants in
  $\cH_{el}$ are renormalized only trivially by rescaling in a RG
  procedure.  This is a well known property of systems consisting of
  an elastic term diagonal in replicas and an interaction term, that
  depends only locally on differences of replica fields. It is often
  refered to as `tilt symmetry' \cite{Schulz+88,Balents+93} although
  here it is generalized to tilt, shear and compression moduli.  As
  the symmetry is fulfilled in the present case only for larger scales,
  elastic constants will be renormalized weakly on small scales. We start our
  description there and the effective constants shall for
  convenience again be denoted by $c_{11}, c_{44}$ and $c_{66}$.
\item[(ii)] The prime measure for residual translational order in the
  system will be the mean squared relative displacements
  \[
  B_{\alpha\beta}(\br)=\overline{\langle
    [u_\alpha(\br)-u_\alpha(\bN)][u_\beta(\br)-u_\beta(\bN)]\rangle}.
  \]
  Their scaling $B_{\alpha\beta}(\br)\sim |\br|^{2\zeta}$ defines the
  roughness exponent $\zeta$ of the lattice. From (a) and (c) follow
  the isotropic relation $B_{xx}(x,y,z)=B_{yy}(y,-x,z)$ and analogues.
\item[(iii)] Symmetries (b) and (c) give
  $B_{xx}^\kappa(\br)=B_{yy}^{\kappa^{-1}}(\br)$ and
  $B_{xy}^\kappa(\br)=-B_{xy}^{\kappa^{-1}}(\br)$, which are relations
  between correlation functions of {\em different} flux line lattices
  which are related by an inverted ratio $\kappa=c_{66}/c_{11}$ of
  elastic constants as indicated by the superscript on
  $B^\kappa_{\alpha\beta}(\br)$.
\end{itemize}

%---------------------------
%---------------------------
\section{Functional renormalization group}
\label{frg}
%---------------------------
%---------------------------
As was motivated in the last Section, in $d<4$ dimensions we have to
deal with infinitely many relevant operators. Therefore, we employ a
{\em functional} renormalization group (FRG) method to treat the
interaction potential $R(\bu)$ between different replicas. In the
past, this technique had been successfully applied to single elastic
objects in random environments \cite{FisherDS86b,Balents+93}. Later
the same method was used to describe lattices of elastic objects like
that of flux lines \cite{Giamarchi+95}. However, the model studied by
FRG for the latter case has two important drawbacks: The effect of the
triangular lattice is neglected in both (i) the elasticity, that is
anisotropic for all physical FL lattices and (ii) the disorder
correlator. The latter reflects the lattice symmetry and any RG flow
will have to preserve it.  In the following we develop a FRG approach
to take into account both effects, which result in new physical
behaviour.
%---------------------------
\subsection{RG equations}
\label{eof}
%---------------------------
We use a standard hard-cutoff RG by integrating out the displacement
field $\bu(\bq)$ with wavevectors $\bq$ in an infinitesimal momentum
shell below the cutoff $\label{mom-shell}
\Lambda > |\bq| > \Lambda/ b \equiv \Lambda e^{-\mbox{\scriptsize d} l}$.
The contributions to the RG equations from rescaling according to
\begin{eqnarray}
\label{rescaling}
\bq&=&\bq'/b \nn\\
\br&=&\br'\,b \nn\\
\bu(\br)&=& \bu'(\br')\,b^\zeta 
\end{eqnarray}
are given by
\begin{eqnarray}
{{\partial T} \over {\partial l}}\left. \right|_{sc} &=& (2-d-2\zeta)T
\label{scaling1}\\
{\partial R \over \partial l}\left. \right|_{sc} &=& (4-d-4\zeta)R(\bu)
        + \zeta \partial_\alpha R(\bu) u_\alpha.
\label{scaling2}
\end{eqnarray}
Since the order of the flux line lattice is expected to be dominated
by the random potential, we have chosen to rescale the temperature in
order to organize the RG analysis of the expected $T=0$ fixed point.
The corresponding Eq. (\ref{scaling1}) is exact, i.e., there will be
no feedback from the random potential to the elastic constants as
mentionend above. From the second term on the rhs of Eq.
(\ref{scaling2}) it is obvious that the periodicity of $R(\bu)$ is not
consistent with a finite roughness exponent $\zeta$. Therefore, we
have to choose $\zeta=0$ to obtain a periodic fixed point function
$R^*(\bu)$. This choice of $\zeta$ automatically implies smaller than
power-law roughness on asymptotic scales {\em if} a fixed point is
found. Still, choosing $\zeta$ to be zero is just a matter of the
fixed point analysis and does not forbid finite values for $\zeta$ on
smaller length scales.

To obtain a fixed point for $R(\bu)$ to order $\epsilon=4-d$, we have
to calculate all terms of second order in $R(\bu)$ which contribute to
Eq. (\ref{scaling2}). To do so, we trace over the fast modes of
$\bu(\br)$. The feedback to the disorder energy can be written in a
cumulant expansion in $\cH_{dis}$. Taking into account the anisotropy
of the elastic kernel, we obtain the RG equation
\begin{eqnarray}
\label{eof_raw}
\partial_l R(\bu) &=&
        \epsilon R(\bu) + \left({1\over 2} \partial_\alpha \partial_\gamma
        R(\bu) \partial_\beta \partial_\delta R(\bu) - \right. \nn \\
&&\hspace*{.6cm} \left. \phantom{{1\over 2}}-
        \partial_\alpha \partial_\gamma R(\bu)
        \partial_\beta \partial_\delta R(\bN)\right)
                \bM^{\alpha \beta,\gamma \delta}
\end{eqnarray}
\begin{eqnarray}
\label{M}
\mbox{with }\quad \bM^{\alpha \beta, \gamma \delta}
&=& \d l^{-1} \int^>_\bq \cG_{\alpha \beta}(\bq) \cG_{\gamma \delta}(-\bq)
\end{eqnarray}
with $\cG_{\alpha \beta}=\cG_L \bP^L_{\alpha \beta} + \cG_T
\bP^T_{\alpha \beta}$.  Here $\int_\bq^>$ denotes integration over the
shell $\Lambda > |\bq| > \Lambda e^{-\mbox{\scriptsize d} l}$ to order $\d l$.  The
symmetric matrix $\bM$ with the greek double index ordered according
to $(xx,xy,yx,yy)$ is computed in appendix
\ref{appendix-eof} as
\begin{equation}
\label{M-computed}
\bM={1 \over 8} 
        \left( \begin{array}{cccc}
        3 I_1 + 2 I_2 & 0 & 0 &  I_1 + 6 I_2 \\
        0 &  I_1 - 2 I_2 & I_1 - 2 I_2& 0 \\
        0 & I_1 - 2 I_2& I_1 - 2 I_2& 0 \\
        I_1 + 6 I_2 & 0 & 0 & 3 I_1 + 2 I_2
        \end{array} \right)
\end{equation}
with integrals
\begin{eqnarray}
\label{Is}
I_1 &\equiv&
      \d l^{-1}  \int_\bq^> \left( \cG^2_T + \cG^2_L \right)=
      {1\over {8\pi^2}}
        {{1+\kappa}\over {c_{44} c_{66}}}  \nn\\
&&\nn\\
I_2 &\equiv&
        \d l^{-1} \int_\bq^> \cG_T \cG_L =\frac{1}{8\pi^2 c_{44}c_{11}}
        \frac{\ln \kappa}{\kappa-1}  \:.
\end{eqnarray}
In the spirit of a consistent $\epsilon$-expansion, the integrals are
evaluated in 4D.

With the shorthand notations $-\Delta\equiv R_{xx}(\bN) =
R_{yy}(\bN)$, $\delta\equiv 1-2I_2/I_1$, and with the rescaling
\begin{equation}                                                        
\label{Rrescaling}
R\equiv\tilde{R} {{2 a^2}\over {I_1}}, \quad \quad
     \Delta=\tilde{\Delta}{{2a^2}\over { I_1}},
\end{equation} 
which makes the parameter $\tilde{\Delta}$ dimensionless, the RG
equation finally becomes
\begin{eqnarray}
\label{eof-final}
\lefteqn{
{\partial \tilde{R}(\bu) \over {\partial l}}=
        \epsilon \tilde{R}(\bu) 
}\hspace{-.1cm}\nn\\
&&   + {a^2 \over 2}  \Big\{
   \tilde{R}_{xx}^2(\bu)+\tilde{R}_{yy}^2(\bu) + 
   2 \tilde{R}_{xy}^2(\bu) + 2 \tilde{\Delta}
        [\tilde{R}_{xx}(\bu) \nn \\
&& + \tilde{R}_{yy}(\bu)]
           -\!{\delta \over 4} ([\tilde{R}_{xx}^2(\bu)
           -\tilde{R}_{yy}^2(\bu)]^2 +
           4\tilde{R}_{xy}^2(\bu))\Big\}.
\end{eqnarray}
It is easily checked that the RG flow given by Eq. (\ref{eof-final})
preserves the symmetries of the correlator $R(\bu)$.  The RG flow
depends on the ratio $\kappa=c_{66}/ c_{11}$ via the anisotropy
parameter
\begin{equation}
\delta=1-2{{\ln \kappa} \over
{\kappa -
\kappa^{-1}}},
\end{equation}
which varies between $0$ and $1$. This dependence on the elastic
constants is inherently related to the anisotropic elasticity and
cannot be eliminated by further rescaling as it is the case for the
more isotropic idealizations of flux line lattices studied in Ref.
\onlinecite{Giamarchi+95}. Below it will be shown that the 
{\em nonuniversality} of the coefficients in the RG
Eq. (\ref{eof-final}) carries through to nonuniversal exponents for
the displacement correlations. To compare with former studies, we
consider two limiting cases of Eq. (\ref{eof-final}): \newline(i)
$c_{11}=c_{66}$ corresponding to $\delta=0$. This is the isotropic
case which former studies were restricted to, mainly for technical
reasons, since this extreme limit is not realized in isotropic
superconductors at low temperatures.  \newline(ii) $c_{11} \gg c_{66}$
corresponding to $\delta=1$. This case is realized physically near the
upper critical field $H_{c2}$ where the lattice becomes very soft with
respect to shear stress. In the very extreme limit $c_{11}\to\infty$
only the transverse propagator is effective in the Hamiltonian and
Eq. (\ref{eof-final}) can be rescaled as to be free of the remaining
elastic constant $c_{66}$.
%---------------------------
\subsection{Numerical solution}
\label{numerical_solution}
%---------------------------
The quantity which will determine the effective propagator of the
displacement field is the renormalized disorder strength $\Delta_l$,
considered as a function of the logarithmic length scale or RG
parameter $l$. Since the RG flow of Eq.  (\ref{eof-final}) cannot be
reduced to that of $\Delta_l$ only, we have to solve the full partial
differential equation for the function $R_l(\bu)$, starting from its
bare value $R_0(\bu)$ given by Eq. (\ref{physicalR}). Even for the
fixed point condition $\partial R(\bu)/\partial l=0$ an analytical
solution is not obvious. Therefore, we treat the equation of flow 
numerically. Rather than solving the fixed point equation, we
obtain the fixed point by integrating Eq. (\ref{eof-final})
numerically from the starting function $R_0(\bu)$. This is necessary
both to exclude non-physical fixed point solutions, which are not
connected by an RG flow to $R_0(\bu)$ and to get additional
information about the behaviour on intermediate length scales, see
Section \ref{three_regimes}.

\begin{figure}
  \resizebox{0.45\textwidth}{!}{\includegraphics{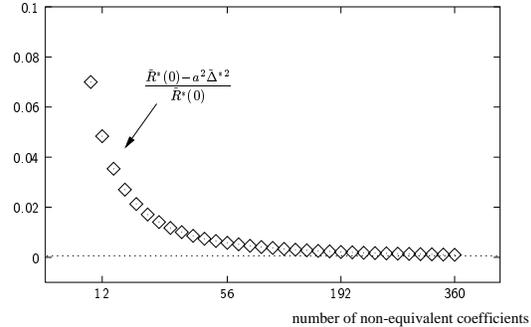}}
    \caption{Numerical accuracy 
      as a function of non-equivalent Fourier coefficients, i.e.
      coefficients that are not related by a point group symmetry. Our
      extrapolation method yields for the test quantity $(\epsilon
      \tilde{R}^*(\bN)+a^2 (\tilde{\Delta}^*)^2)/\epsilon\tilde
      R^*(\bN)=6\cdot 10^{-5}$ instead of the theoretically expected
      value of zero.  This gives the precision for $\tilde\Delta^*$
      mentioned in the text.} \label{check}
\end{figure}

As mentioned above, the lattice symmetry of the initial function
$R_0(\bu)$ is preserved by the RG flow. Technically, this allows
easily for a numerically stable and effective iteration of the RG flow
to large $l$ by rewriting $\tilde R(\bu)$ as the Fourier series
\begin{eqnarray}
\label{ansatz}
\tilde{R}(\bu)&=&\sum_{\bG} \tilde{R}_\bG \cos(\bG\bu)
\end{eqnarray}
with identical coefficients for reciprocal lattice vectors related by
a pointgroup symmetry.  The partial differential equation for $\tilde
R(\bu)$ becomes with this ansatz an infinite set of coupled ordinary
differential equations for the coefficients $\{\tilde R_\bG\}$. Of
course, the numerical integration has been restricted to a finite set
of coefficients $\tilde R_\bG$ by choosing a cutoff $G_c$ so that
$|\bG|<G_c$. The accuracy of the computation  depends on
$G_c$. To obtain numerical results for $G_c\to\infty$, we employ a
routine typically used for finite size scaling of numerical data to
get sufficient precision for $\partial_{xx}
\tilde{R}^*(\bN)=\partial_{yy}
\tilde{R}^*(\bN)=-\tilde{\Delta}^*$, the quantity entering further
calculations.  How reliable the numerical results indeed are, can be
estimated from the exact relation for the fixed point correlator
$\epsilon \tilde{R}^*(\bN)+a^2 (\tilde{\Delta}^*)^2=0$ that stems from
Eq. (\ref{eof-final}), see Figure \ref{check}.  The accuracy of 4
significant digits of $\tilde{\Delta}^*$ can be assured. 

The variable parameters of the initial function $\tilde R_0(\bu)$ are the
number of non-zero coefficients $\tilde R_\bG$, which is fixed by the
value of $\xi$ in Eq. (\ref{physicalR}), and the magnitude of the
coefficients $\tilde R_\bG$. Notice that {\em all} $\tilde R_\bG$ are driven
to non-zero values by the RG flow even if their bare values were
choosen to be zero at the beginning.  Whereas the choice of $\xi$ in
relation to $a$ is important for the existence of an intermediate
length scale regime with a non-trivial finite $\zeta$ (see Section
\ref{three_regimes}), the magnitude of the $\tilde R_\bG$ determines just the
crossover length scale to the asymptotic fixed point function
$\tilde R^*(\bu)$, which itself is universal.  This universal behaviour has
been confirmed numerically for a very wide range 
of starting values. One can write the  condition for finding
convergence to the fixed point as a lower limit for the positional
correlation length $L_a \gtrsim 10^2\Lambda^{-1}$. $\Lambda$ is the
small lengthscale cutoff and of order $\lambda$, the London
penetration length. This condition can now be compared to
\[
L_a\gtrsim C(a_c^2+\lambda^2)^{1/2}
\]
which is the weak disorder condition necessary for the system to be
stable against the formation of dislocation loops \cite{Kierfeld+97}.
$C$ is a constant of order one and $a_c$ the core radius of a
dislocation.  One recognizes that the bassin of attraction of the
fixed point covers well the range of validity of our dislocation-free
description.

\begin{figure}
  \resizebox{0.45\textwidth}{!}{
    \includegraphics{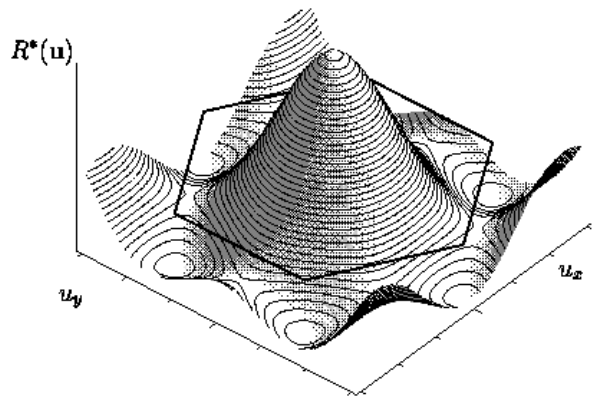}
  }
  \caption{Fixed point correlator $\tilde{R^*}(\bu)$. The hexagon represents a Wigner-Seitz cell of the 
    triangular lattice.}
  \label{fixpointfct}       
\end{figure}

\begin{figure}
  \resizebox{0.45\textwidth}{!}{ \includegraphics{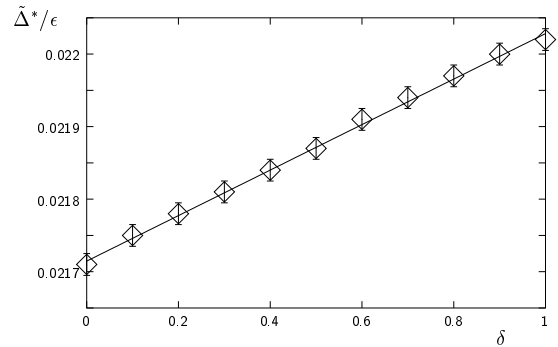}
    } \caption{Variation of $\tilde\Delta^*=-\tilde R_{xx}^*(\bN)=-\tilde R_{yy}^*(\bN)$
    with anisotropy $\delta=1-2\ln \kappa/(\kappa-\kappa^{-1})$.}
    \label{Delta_vs_delta}
\end{figure}

In Figures \ref{fixpointfct}, \ref{correlator_l_20} the fixed point
function and its second radial derivative are shown.  Fig.
\ref{correlator_l_20} shows clearly the existence of a non-analytic
cusp at the lattice sites. Dependency of $R^*(\bu)$ on the anisotropy is
obvious from the numerical results for $\tilde \Delta^*$ as shown in
Fig. \ref{Delta_vs_delta}.

%---------------------------
%---------------------------
\section{Correlation functions}
\label{correlations}
%---------------------------
%---------------------------

%%---------------------------
\subsection{Length crossover effects}
\label{three_regimes}
%---------------------------
With the solution of the RG equation at hand, we calculate the
disorder and thermal averaged squared relative displacement as a
measure of translational order.  It is related to the basic correlator of
the replica fields 
\[
\langle u_\alpha^a(\bq) u_\beta^b(\bq') \rangle=T\cG_{\alpha\beta}^{ab}(\bq)\,
(2\pi)^d\delta(\bq+\bq')
\]
by 
\begin{eqnarray}
\label{B-Fourier}
B_{\alpha \beta}(\br-\br')&=&
\overline{\langle [u_{\alpha}(\br)-u_{\alpha}(\br')]
                [u_{\beta}(\br)-u_{\beta}(\br')]\rangle}\nn\\
&=& 2 \!\lim_{n\to 0} \int\limits_{\bq}\!\!T\cG_{\alpha \beta}^{a a}(\bq) 
        \left(1\!-\!\cos \bq(\br-\br')\right).
\end{eqnarray}
If we denote the correlator for a system with a renormalized
Hamiltonian $\cH_l^n$ by $\cG_l$ and if we allow temperature to flow
according to Eq. (\ref{scaling1}), the scaling relation
$T\cG(\bq)=T_l\cG_l(b\bq)b^{2\zeta+d}$ holds for all $\bq$ with
$|\bq|\le \Lambda/b$. To obtain to order $\epsilon$ the exact
propagator $\cG(\bq)$ for fixed $\bq$, we choose $b=\Lambda/|\bq|$.
This allows to employ a harmonic approximation to calculate $\cG_l$
since the coupling of modes with momenta between $|\bq|$ and $\Lambda$
has been taken into account already by the renormalization of
$\cH_l^n$. Thus we obtain the propagator
\begin{eqnarray}
\label{propagator}
T\cG_{\alpha\beta}^{aa}(\bq)
&=&
\left\{ \left({\Lambda \over q}\right)^{d-4}
\!\!\!\!\!\!\Delta_l \left(\cG_L^2(\bq)\bP^L_{\alpha\beta}+
  \cG_T^2(\bq)\bP^T_{\alpha\beta}\right)\right. 
\nn\\ 
&&+T\left(\cG_L(\bq)\bP^L_{\alpha\beta}+
  \cG_T(\bq)\bP^T_{\alpha\beta}\right)\Bigg\}
\end{eqnarray}
with $l=\ln{\Lambda/q}$. The term $\sim T$ is the mere
thermal propagator. It does not effect roughness in $d > 2$ below the
melting transition and is thus negligible against the disorder term
$\sim \Delta_l$ which is more strongly divergent for $|\bq| \to 0$.

\begin{figure}
  \resizebox{0.45\textwidth}{!}{
    \includegraphics{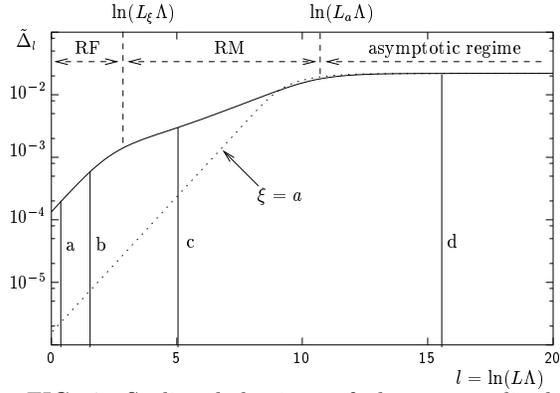} } \caption{Scaling
    behaviour of the renormalized disorder strength $\Delta_l$ in the
    three different length scale regimes (solid curve). The random
    manifold (RM) regime disappears (dashed curve) if the bare $R_0(\bu)$
    has non-vanishing Fourier coefficients for the lowest order
    hexagon only, corresponding to the case $a=\xi$ [see
    Eq. (\ref{physicalR})].}  \label{three_regimes_15_2}
\end{figure}

\begin{figure}
  \resizebox{0.45\textwidth}{!}{ \includegraphics{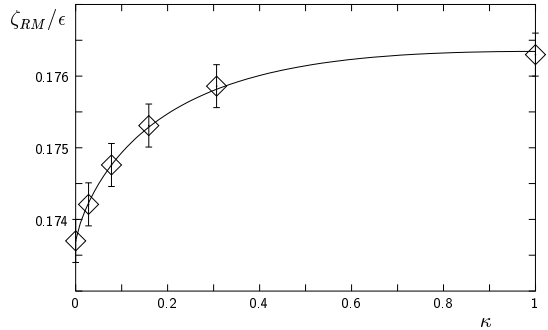} }
    \caption{The roughness exponent $\zeta_{\rm RM}$ is nonuniversal
    in the RM regime. } \label{zeta_vs_kappa}
\end{figure}

Before we discuss the asymptotic displacement correlations in the next
Section, we want to exploit the behaviour of $B_{\alpha\beta}(\br)$ on
intermediate length scales. According to the definition of the
roughness exponent $\zeta$ by $B(\br)\sim |\br|^{2\zeta}$ the scaling
of the renormalized disorder strength $\Delta_l$ determines $\zeta$.
It is easily observed from Eq. (\ref{propagator}) that necessarily
$\ln \Delta_l\sim 2\zeta l$ before it reaches its fixed point
$\Delta^*$ corresponding to $\zeta=0$. The RG flow of $\ln \Delta_l$
as a function of $l$ is plotted in Fig. \ref{three_regimes_15_2}.
Two qualitative different types of behaviour can be observed,
depending on the bare form of the correlator $R_0(\bu)$: (i) If
$a=\xi$ only the lowest order harmonics of $R_0(\bu)$ are non-zero and
two different scaling regimes with a sharp crossover emerge. The first
one is called random force (RF) regime since $\zeta=\zeta_{\rm
  RF}=\epsilon/2$ as predicted by Larkin's perturbative approach for a
random force model \cite{Larkin70}, which becomes applicable if
$B(\br) \lesssim \xi^2$. The last scaling regime is the asymptotic (A)
one with logarithmic roughness. (ii) If $a>\xi$ also higher order
Fourier coefficients are non-zero and a new scaling regime with a
non-trivial value of $\zeta$ appears between the RF and A regime. It
is called random manifold (RM) regime since here the roughness is
expected to be determined by the wandering of lines that do not yet
compete with each other for disorder energy minima, i.e., $B(\br)
\lesssim a^2$. But metastability is already important on these length
scales leading to a new roughness exponent $\zeta_{\rm RM}$. One of
the central results of our analysis is that $\zeta_{\rm RM}$ is {\em
  nonuniversal}. It depends on the ratio $\kappa$ of elastic
constants and varies between $0.1737\epsilon$ and $0.1763\epsilon$ as
shown in Fig. \ref{zeta_vs_kappa}.
The only result for $\zeta_{\rm RM}$, which has been available so far, is
based on a Flory type argument for the flux line lattice in that
regime \cite{Kardar87,Nattermann87}. It is given by
$\zeta_{RM}=\frac{4-d}{4+N}=\epsilon/6 \backsimeq 0.167\epsilon$ ($N$
is the number of components of the displacement field, here $N=2$.)
and supposed to be a lower bound, in agreement with our findings. We
expect that the range of variation of the nonuniversal $\zeta_{RM}$
is larger in real three dimensions than an epsilon expansion to first
order can reveal. This is because in the propagator the effect of
variing $c_{11}, c_{66}$ is suppressed in an expansion around 4D by
the more heavily weighted $c_{44}\bq_z^2$-term.
 
\begin{figure}[h]
  \resizebox{0.45\textwidth}{!}{
    \includegraphics{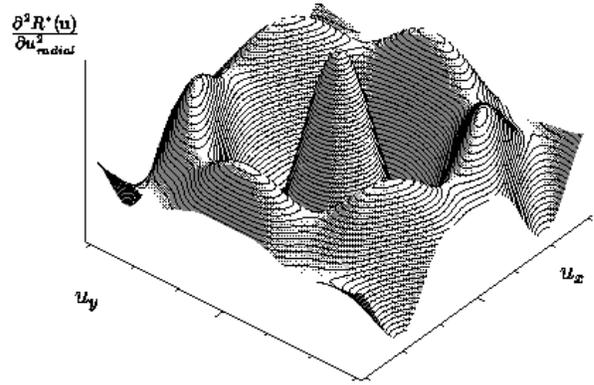} } \caption{Second radial
    derivative $\partial_{|\bu|}^2R(\bu)$ at the initial position a
    ($l=0$) shown in Fig. \ref{three_regimes_15_2}.}
    \label{correlator_l_001}
\end{figure}

\begin{figure}[h]
  \resizebox{0.45\textwidth}{!}{ \includegraphics{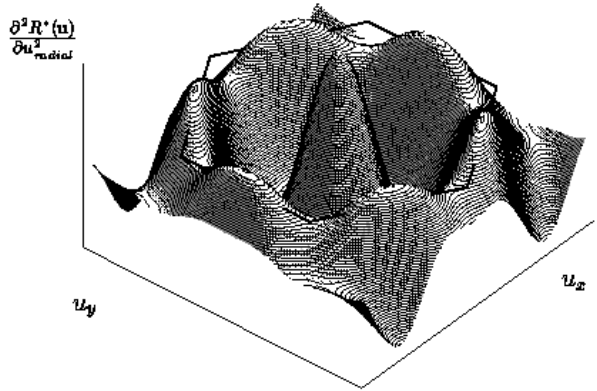}
    } \caption{Random force regime: $\partial_{|\bu|}^2R(\bu)$ at
    position b in Fig. \ref{three_regimes_15_2} obtained by
    integrating numerically the RG Eq. (\ref{eof-final}).}
    \label{correlator_l_2}
\end{figure}

\begin{figure}[h]
  \resizebox{0.45\textwidth}{!}{
    \includegraphics{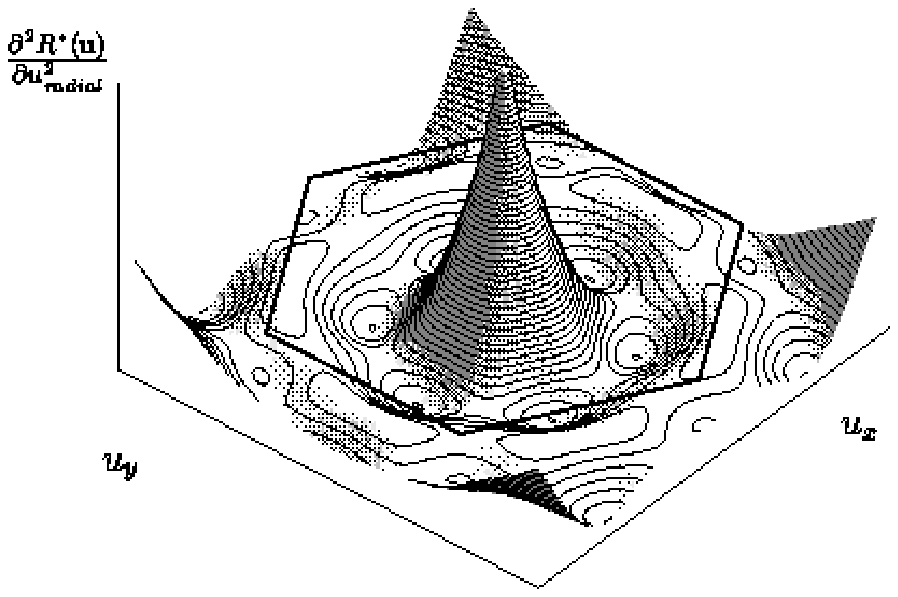}
  }
  \caption{Random manifold regime: $\partial_{|\bu|}^2R(\bu)$ at
    position c in Fig. \ref{three_regimes_15_2}.}
  \label{correlator_l_8}       
\end{figure}

\begin{figure}[h]
  \resizebox{0.45\textwidth}{!}{
    \includegraphics{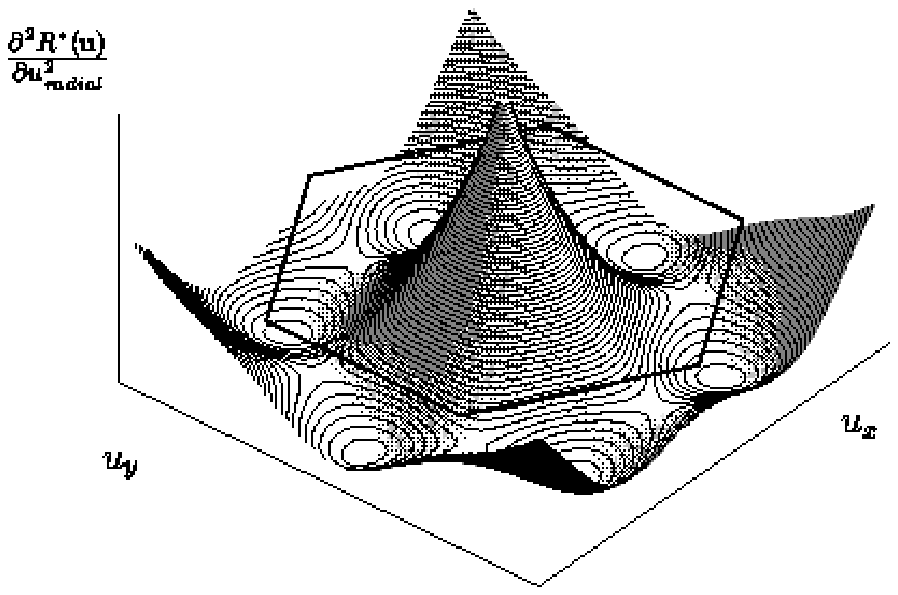}
  }
  \caption{Asymptotic fixed point: $\partial_{|\bu|}^2R(\bu)$ at
    position d in Fig. \ref{three_regimes_15_2}.}
  \label{correlator_l_20}       
\end{figure}

Next we calculate the crossover length scales. These are denoted by
$L_\xi$ and $L_a$ for the crossover RF -- RM and RM -- A,
respectively. $L_\xi$ can be obtained from Larkin's perturbative
analysis, which yields for the displacement correlations
\begin{eqnarray}
\lefteqn{
B(\br)=\overline{\langle[\bu(\bx,z)-\bu(\bN,0)]^2\rangle}\backsimeq
\frac{B}{\Phi_0}\frac{f^2_{pin}n_{imp}\xi^2}{\pi c_{44}^{1/2}}
}\hspace{.2cm}\nn\\ &&\times \left\{ c_{11}^{-3/2}(\bx^2+z_l^2)^{1/2}
+ c_{66}^{-3/2}(\bx^2+z_t^2)^{1/2}\right\}.
\end{eqnarray} 
Here we have introduced the rescaled z-coordinates $z_l=z
\sqrt{c_{11}/ c_{44}}$, $z_t=z \sqrt{c_{66}/
c_{44}}$. The anisotropic crossover or Larkin length scale is
determined by the conditions $B(\bN,z=L_\xi^z)\simeq \xi^2$ and
$B(|\bx|=L_\xi^\bx,0)\simeq\xi^2$, thus giving
\begin{eqnarray}
\label{Larkinlength}
L_\xi^z & \backsimeq &
\frac{\pi \Phi_0}{B f_{pin}^2n_{imp}}
\frac{c_{11}c_{44}c_{66}}{c_{11}+c_{66}}\nn\\
L_\xi^\bx & \backsimeq & \frac{\pi \Phi_0}{B f_{pin}^2n_{imp}}
\frac{c_{44}^{1/2}(c_{11}c_{66})^{3/2}}
{c_{11}^{3/2}+c_{66}^{3/2}}.
\end{eqnarray}
Based on the condition for the mean displacements of lines at the
crossover scales, the second crossover length scales $L_a^z$,
$L_a^\bx$ should be related to the Larkin length scales by
$L_a^{(z,\bx)}\simeq (a/\xi)^{1/\zeta_{\rm RM}}L_\xi^{(z,\bx)}$.  To
check this relation we have determined the crossover scales
numerically from the RG flow of $\Delta_l$ for different ratios
$a/\xi$.  The estimates for $\zeta_{RM}$ thus obtained are consistent with the exact
values shown in Fig. \ref{zeta_vs_kappa}.  Of course, due to the
finite extent of the crossover regions, errors are much too large as
to take this as a measurement of $\zeta_{RM}$; however, it clearly
confirms the physical picture of the crossover lengthscales.  Turning
to the RG flow of the whole function $R_l(\bu)$, one expects that the
appearance of metastability on the Larkin scale should be reflected in
a change of the functional form of $R_l(\bu)$, too. Indeed, the
existence of a cusp in the RM and A regime can be observed from the
sequence of Figures \ref{correlator_l_001}, \ref{correlator_l_2},
\ref{correlator_l_8} and \ref{correlator_l_20}, which show the second
radial derivative of $R_l(\bu)$ in the different scaling regimes, see
Fig. \ref{three_regimes_15_2}. From this observation we expect that
the fourth derivatives
$\partial^4_{u_x}\tilde{R}(\bN)=\partial^4_{u_y}\tilde{R}(\bN)
\equiv\tilde{R}^{(4)}(\bN)$ diverge at the scale given by the Larkin
length.  The scale $l_c=\ln(L_c \Lambda)$ at which the cusp appears
can be extracted from the RG Eq. (\ref{eof-final}) as pointed out in
Ref. \onlinecite{Gorokhov+99} for a different RG equation.  Using the
relations $\partial^4_x\tilde{R}(\bN)=\partial^4_x\tilde{R}(\bN)=
3\partial^2_x\partial^2_y\tilde{R}(\bN)$ amongst derivatives of the
correlator at the lattice sites, we get as equation of flow for the fourth
derivative $\tilde{R}^{(4)}(\bN)$ at the lattice sites
\begin{equation}
\partial_l\tilde{R}^{(4)}(\bN)=\epsilon \tilde{R}^{(4)}(\bN) +
\frac{a^2}{3}(10-\delta)
\left[\tilde{R}^{(4)}(\bN)\right]^2
.\end{equation}
This equation can be easily integrated and yields in the weak disorder
limit for the length scale where first $\tilde{R}^{4}(\bN)=\infty$ the result
\begin{equation}
\label{Lcrossover}
L_c=\frac{3}{a^2\Lambda (10-\delta)} \frac{1}{\tilde R^{(4)}_0(\bN)}.
\end{equation}
Using relation (\ref{physicalR}), (\ref{Rrescaling}) the bare value of
$\tilde{R}_0^{(4)}$ can be expressed in physical quantities and we get
for $\xi\ll a$
\begin{equation}
L_c\backsimeq\frac{1}{2}\frac{1}{\rho_0 f_{pin}^2n_{imp}}
\frac{c_{44}c_{11}c_{66}}{c_{11}+c_{66}}.
\end{equation}
Comparison of this result to the Larkin length of
Eq. (\ref{Larkinlength}) shows that the correlator $R_l(\bu)$ becomes
indeed non-analytic at the crossover to the RM regime.
%
%---------------------------
\subsection{Asymptotic displacement correlations}
\label{asymptotic_displacement_correlations}
%---------------------------
We now study the correlations in regime A in detail, starting with the
displacement correlations. Beyond the second crossover scale $L_a$ the
larger-$q$ modes governing both the RF and RM regime become
unimportant and we may benefit from the fact that $\Delta_l\to
\Delta^*$ in the asymptotic regime. Therefore, the only relevant
disorder part of the correlator in Eq. (\ref{propagator}) behaves like
$T\cG_{\alpha\beta}^{aa}(\bq) \sim \Delta^* q^{-d}$ leading to
logarithmic roughness. It is interesting to note that the amplitude of
the displacement correlations $B_{\alpha\beta}(\br)$ does not depend
on the disorder strength, but is proportional to the fixed point value
$\tilde\Delta^*(\kappa)$, whose dependency on $\kappa$ is shown in
Fig. \ref{Delta_vs_delta}.  The correlation function
$B_{\alpha\beta}(\br)$ can now be calculated with explicit
consideration of the anisotropic elasticity using Eqs.
(\ref{B-Fourier}), (\ref{propagator}).  With the abbreviations $z_l=z
\sqrt{c_{11}/c_{44}}$, $z_t=z \sqrt{c_{66}/c_{44}}$ and
$h(t)=t^{-2}\ln(1+t^2)$ we find in regime A
\begin{eqnarray}
\label{asymptotic-B}
B_{xx}(\br)&=&\frac{\tilde{\Delta}^*(\kappa)a^2}{1+\kappa}\left\{
\ln\!\left(\frac{\bx^2+z_{t}^2}{L_a^2}\right)
+\kappa \ln\!\left(\frac{\bx^2+z_{l}^2}{L_a^2}\right)
\right.\nn\\
&&+\left.\frac{y^2-x^2}{\bx^2}\left[1-\kappa-h\!\left(\frac{|\bx|}{z_t}\!\right)+\kappa \, 
h\!\left(\frac{|\bx|}{z_l}\!\right)\right]\right\},\nn\\[1\bs]
B_{yy}(\br)&=&\frac{\tilde{\Delta}^*(\kappa)a^2}{1+\kappa}\left\{
\ln\!\left(\frac{\bx^2+z_{t}^2}{L_a^2}\right)
+\kappa \ln\!\left(\frac{\bx^2+z_{l}^2}{L_a^2}\right)
\right.\nn\\
&&+\left.\frac{x^2-y^2}{\bx^2}\left[1-\kappa-h\!\left(\frac{|\bx|}{z_t}\!\right)+\kappa \, 
h\!\left(\frac{|\bx|}{z_l}\!\right)\right]\right\},\nn\\[1\bs]
B_{xy}(\br)&=&\frac{2\tilde{\Delta}^*(\kappa)a^2}{1+\kappa}\frac{xy}{\bx^2}
\left\{\kappa-1-\kappa \, h\!\left(\frac{|\bx|}
{z_l} \right)+h\!\left(\frac{|\bx|}
{z_t} \right)\right\}\nn\\
\end{eqnarray}
and thus
\begin{equation}\begin{split}
&\overline{\langle[\bu(\br)-\bu(\bN)]^2\rangle}=
B_{xx}(\br)+B_{yy}(\br)\\ 
&=\frac{2\,\tilde{\Delta}^*(\kappa)a^2}{1+\kappa}\left\{
\ln\!\left(\frac{\bx^2+z_{t}^2}{L_a^2}\right)
+\kappa \ln\!\left(\frac{\bx^2+z_{l}^2}{L_a^2}\right)
\right\}.
\end{split}\end{equation}
The details of the Fourier transformation are given in appendix
\ref{fourier_transform}. There the calculation retaining all unbounded
terms for large $|\bx|,z$ is performed in 4D since $\tilde{\Delta}^*$
is $\cO(\epsilon)$. For the reason mentioned in Section \ref{symmetries} (b)
$B_{yy}(\br)$ can be obtained from $B_{xx}(\br)$ by $\kappa
\rightarrow \kappa^{-1}$. The relative line displacement
$B_{\alpha\beta}(\br)$ grows only logarithmically due to the
restriction of available configurations by adjacent lines. The
`geometric' coefficients of the logarithms reflect the elastic
anisotropy.  The result reminds of the findings for the analogous 2D
flux line lattice by Carpentier and Le Doussal, where there is also a
nonuniversal prefactor of logarithmic roughness that depends on the
elastic details of the system \cite{Carpentier+97}.

To compare our results with former results by Giamarchi and Le Doussal (GD)
\cite{Giamarchi+94,Giamarchi+95} in 3D we consider two limiting cases. First, 
in the limit $z\to 0$ the relative displacement reduces to
\begin{eqnarray*}
B_{xx}&=&\tilde{\Delta}^*(\kappa)a^2
 \left\{\ln\!\left(\frac{\bx^2}{L_a^2}\right)+
\frac{x^2-y^2}{\bx^2}\frac{\kappa-1}{\kappa+1}\right\}\\
B_{xy}&=&2\tilde{\Delta}^*(\kappa)a^2\frac{xy}{\bx^2}\frac{\kappa-1}{\kappa+1}
\end{eqnarray*}
This result is identical in form with the result of GD obtained from a
variational ansatz treatment for the triangular lattice.  However,
this ansatz yields a value for $\tilde\Delta^*$ which is independent
of $\kappa$ and differs from ours by about 15\%. The origin of this
deviation is the inexact treatment of fluctuations by the variational
method.  The renormalization group approach of GD is restricted to an
idealized scalar field model of a flux line lattice with isotropic
elasticity. To determine the influence of the triangular lattice
symmetry on $\tilde \Delta^*$ compared to the scalar model, we
consider now the isotropic case $c_{11}=c_{66}$ corresponding to
$\kappa=1$. With the rescaled coordinate
$\br'=(\bx,\sqrt{c_{11}/c_{44}}z)$ we obtain
\begin{equation*}
B_{xx}=B_{yy}=2\tilde{\Delta}^*(1)a^2\ln(|\br'|/L_a),\quad B_{xy}=0
\end{equation*}
with $\tilde \Delta^*(1)=0.0217\epsilon$. This has to be compared to 
the result $\tilde \Delta^*(1)=\epsilon/36\simeq 0.028\epsilon$ 
for the scalar model.

%---------------------------
\subsection{Order parameters}
\label{order_parameters}
%---------------------------

\begin{figure}
  \resizebox{0.45\textwidth}{!}{
    \includegraphics{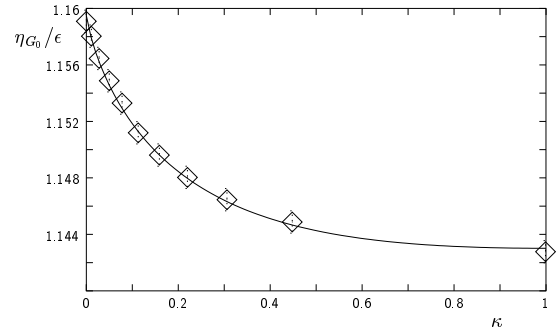}
  }
  \caption{Dependency of the exponent $\eta_{\bG_0}$ on the ratio
    $\kappa=c_{66}/c_{11}$ of elastic constants.}
  \label{eta_vs_kappa}       
\end{figure}

Now we study the translational order parameter $\Psi_\bG(\br)\equiv
e^{i\bG\bu(\br)}$ to measure the remaining translational order in the
impure system. The pair correlation function is 
\begin{equation}
\label{Scorr}
C_{\bG}(\br)=\overline{\langle\Psi_\bG(\br)\Psi^*_\bG(\bN)\rangle}=
\overline{\langle e^{i \bG(\bu(\br)-\bu(\bN))}\rangle}.
\end{equation}
It is a preferred measure, as its scaling behaviour determines the
intensity of the reflection pattern obtained in neutron scattering,
see below.  $C_\bG(\br)$ is often referred to as translational order
correlation or simply `translational order'.  Being exact to order
$\epsilon$, the average can be raised to the exponent as if $\bu(\br)$
was Gaussian distributed \cite{phd:Emig}.  We thus can obtain from Eq.
(\ref{asymptotic-B}) the result for the asymptotic scaling regime,
\begin{eqnarray}
\label{transl-order-result}
C_\bG(\br)   &\propto&   g_\bG L_{a}^{\eta_\bG(\kappa)}(\bx^2+z_{t}^2)^
{-\frac{\eta_\bG(\kappa)}{2(1+\kappa)}}\nn \\
&&\times(\bx^2+z_{l}^2)^{-\frac{\eta_\bG(\kappa)}{2(1+1/\kappa)}}
\end{eqnarray}
with exponent
\begin{equation}
\label{eta}
\eta_\bG(\kappa)=\tilde{\Delta}^*(\kappa)(aG)^2
\end{equation}
and the geometrical factor
\begin{eqnarray}
  g_\bG  &=&  \exp\left[\frac{\tilde{\Delta}^*(\kappa)(aG)^2}{1+\kappa}\left(
      (\hat\bx\hat\bG)^2-\frac{1}{2}\right)\! \, \right.\nn\\  
    &&\, \left.\times
    \left\{\!\left(\!1- h\!\left(\frac{|\bx|}{z_t}\right)\!\!\right)
      -\kappa\left(\!1-h\!\left(\frac{|\bx|}{z_l}\right)\!\!\right)\!\right\}
         \right].
\end{eqnarray}
Our main result on the translational order of the flux line lattice
consists in the decay of order with a {\em nonuniversal} exponent
$\eta_\bG(\kappa)$. Its dependency on $\kappa$ is obtained from Eq.
(\ref{eta}) and is shown in Fig. \ref{eta_vs_kappa} for one of the
smallest reciprocal lattice vectors $\bG=\bG_0$ with
$|\bG_0|=4\pi/(a\sqrt{3})$.  The full decay exponent in Eq.
(\ref{transl-order-result}) is modified by additional
$\kappa$-dependent factors due to the anisotropic elasticity leading
to different decay of contributions from transversal and longitudinal
modes.  The nonuniversal behaviour is at variance with the
variational ansatz results and conjectures in Refs.
\onlinecite{Giamarchi+94,Giamarchi+95}.

Let us consider the two limiting cases from above: For $z \rightarrow
0$ the result simplifies to
\begin{eqnarray}
C_\bG(\bx)&\propto& (|\bx|/L_{a})^{-\eta_\bG(\kappa)}\nn\\
&\times&\exp\left\{\tilde{\Delta}^*(\kappa)(aG)^2\frac{1-\kappa}
{1+\kappa}
\left((\hat\bx\hat\bG)^2-\frac{1}{2}\right)\right\},
\end{eqnarray}
which again coincides modulo the difference in the coefficent with the
result found by GD for this limit. For $\kappa=1$, the translational
order decay is isotropic according to
\[
C_\bG(\br')\propto
L_{a}^{\eta_\bG(1)}{|\br'|}^{-\eta_\bG(1)}
\]
with the rescaled coordinate $\br'$ defined above. This isotropic
limit can be used to demonstrate clearly that the triangular flux line
lattice considered here and the scalar model do not belong to the same
universality class. Whereas we obtain $\eta_\bG(1)=1.14\epsilon$ for
the triangular lattice, the result in the RG approach for the scalar
model -- which corresponds to a square lattice -- is
$\eta_\bG(1)=\pi^2/9\epsilon=1.10\epsilon$.

Another order parameter of interest is the {\em positional glass}
correlation function suggested by spin glass theory \cite{Binder+86}
\begin{equation}
S_{PG}(\bG,\br)=\overline{\left|\langle \Psi_\bG(\br)\Psi^*_\bG(\bN)
\rangle\right|^2}.
\end{equation}
It measures the thermal fluctuations of the flux lines around their
disordered ground state.  We calculate $S_{PG}(\bG,\br)$ within our
framework from the hamiltonian renormalized up to scale $|\br|=\Lambda^{-1}
e^{l}$.  First order perturbation theory in the pinning energy gives
the corrections to the mere thermal result to first order in
$\epsilon$.  We get
\begin{equation}
\label{SPG}
S_{PG}(\bG,\br) \backsimeq 
S_{PG}^0(\bG,\br) \left\{1+\epsilon \sum_{m\ge 1}c_m \left(
{T\over r^2}\right)^{2m}\right\},
\end{equation}
where $S_{PG}^0(\bG,\br)$ denotes the correlation function for the
pure system with thermal fluctuations only and $c_m$ are numerical
coefficients. It is finite for $r \rightarrow
\infty$ and reduced with respect to unity merely by the standard
Debye-Waller factor. As the order $\epsilon$ corrections decay they
can surely not compensate the leading constant term and make up for a
more than powerlaw decay of the whole correlation function. This
provides signature of a positional glass to order $\epsilon$.

Widely discussed, however not completely resolved is the question, if
there exists a {\em phase coherent vortex glass} state in impure
type-II superconductors.  A finite value of the correlation function
\begin{equation}
C_{V\!G}(\br)=\overline{\left|\langle\Psi(\br)\Psi^*(\bN)\rangle\right|^2}
\end{equation}
for large $|\br|$ is proposed to identify such a phase coherent vortex
glass \cite{FisherMPA89,Fisher+91}.  Here $\Psi(\br)$ is the
Ginzburg-Landau complex order parameter.  It can be decomposed in
amplitude and phasefactor, $\Psi=|\Psi|e^{i[\phi_0+\delta\phi]}$, with
groundstate phase $\phi_0$ and phase fluctuations $\delta\phi$.  In the
London limit the amplitude is constant outside the vortices and
$C_{V\!G}$ becomes
\begin{equation}
C_{V\!G}(\br)=|\Psi|^2\overline{|\langle e^{i[\delta\phi(\br)-\delta\phi(\bN)]}
\rangle |^2}.
\end{equation}
Since the phase fluctuations are topologically constrained by the
vortex positions, the distortions of the vortex lattice can destroy
phase coherence. More concrete, phase fluctuations and vortex
displacements are related by \cite{Moore92}
\begin{equation}
\nabla^2 \delta\phi(\br)={{2\pi}\over a}\left(\nabla_\perp \times \bu(\br)\right).
\end{equation}
This relation allows for the calculation of $C_{V\!G}(\br)$ in the
framework of the elastic description of the flux line lattice.  We use
perturbation theory for $\cH_{dis}$ with fluctuations on scales below
$|\br|$ having renormalized this perturbation.  Then corrections to
the mere thermal average $\langle\ldots\rangle_0$ are given to first
order in $\epsilon$ by
\begin{equation}\begin{split}
\label{CVG}
& C_{V\!G}(\br)=e^{-\langle [\delta\phi(\br)-\delta\phi(\bN)]^2\rangle_0}
\times \Big\{ 1 \\
& +{1\over T^2}\sum_\bG \hat R^*_\bG \int_{\bar\br}
\left[\cosh(2F(\br,\bar\br))-4\cosh(F(\br,\bar\br))\right]\Big\}.
\end{split}\end{equation}
Here $\hat R_\bG^*=R_\bG^* \exp(-G_\alpha G_\beta\langle
u_\alpha(\bN)u_\beta(\bN)\rangle_0) \sim\epsilon$ are the reduced
Fourier coefficients of the random energy correlator of Eq.
(\ref{correlator}) and
$F(\br,\bar\br)=\Gamma(\br-\bar\br)-\Gamma(-\bar\br)$ with
$\Gamma(\br)=\langle \bG\bu(\br)\delta\phi(\bN)\rangle_0$.  The
exponential factor of $\hat R_\bG^*$ is finite in $d>2$.  For the
disorder induced correction in Eq. (\ref{CVG}) we focus on the
isotropic limit with $c_{44}=c_{66}$ and get
\begin{equation}
\Gamma(\br)=
{T\over{4\pi a^2 c_{44}}}\frac{G_y x -G_x y}{r^2}
\left(1-{2\over{\Lambda r}}J_1(\Lambda r)\right),
\end{equation}
where $J_1(x)$ is the Besselfunction of first kind. Upon expansion of
the $\cosh$-terms, a careful investigation of the behaviour for large
$|\br|$ of the remaining integrals gives to order $\epsilon$
\begin{equation}\begin{split}
&C_{V\!G}(\br)\simeq e^{-\langle[\delta\phi(\br)-\delta\phi(\bN)]^2
\rangle_0}\\
&\times\left\{1+\epsilon\left[d_1 T^2\ln(r\Lambda)+d_2+
\sum_{m\ge0}c_mT^2\left({T\over r}\right)^{2m}\right]\right\}.
\end{split}\end{equation}
The $d_i, c_i$ are again numerical coefficients, yet different
from the ones in Eq. (\ref{SPG}).
The exponential factor of Eq. (\ref{CVG}) is determined by mere
thermal fluctuations, which read
\begin{equation}\begin{split}
\langle[\delta\phi(\br)&-\delta\phi(\bN)]^2\rangle_0\\
&={{8\pi^2}\over{a^4}}\,T\int_\bq {q_\perp^2 \over q^4}
    \cG^T(\bq)[1-\cos(\bq\br)]\sim {1\over\epsilon}|\br|^\epsilon.
\end{split}\end{equation}
Corrections to order $\epsilon$ can thus not compete against the
exponential decay of $C_{V\!G}(\br)$ originating from strong thermal
fluctuations. Thus we conclude that to order $\epsilon$ there is no
phase coherent vortex glass. Whether this result is valid to higher
orders in $\epsilon$ and thus in 3D remains however unclear within the
present analysis. Dorsey et al. \cite{Dorsey+92} indeed found a vortex
glass transition in $6-\epsilon$ dimensions starting from a
Ginzburg-Landau Hamiltonian. If this transition exists down to
three dimensions is not clear.
%---------------------------
\subsection{Experimental implications}
\label{experiment}
%---------------------------
There exists a considerable number of recent experimental studies of
the structure of flux line lattices in high temperature
superconductors.  Neutron diffraction studies provide strong
evidence for the proposed Bragg glass with quasi-longrange order
\cite{Cubitt+93,Zeldov+95,Soibel+00}.  More recently, magnetic decoration
studies on BSCCO showed huge dislocation free regions containing up to
$10^5$ flux lines \cite{Kim+99}. The last result strongly supports our
elastic description, which neglects dislocations.

Comparison with neutron diffraction experiments shows the preferred
role of the translational order correlation $C_{\bG}(\br)$.  The
diverging dependency of the cross section in a neutron scattering
experiment is given by
\begin{equation}
\sigma(\bk) \propto \overline{\langle \rho_\bk \: 
\rho_{-\bk} \rangle}
\propto \tilde S(\bk),
\end{equation}
where $\bk=(\bk_\perp,k_z)$ is the difference between in- and outgoing
wavevectors and $\rho_\bk$ the Fourier transform of the flux line
density in Eq. (\ref{density}). The structure factor $\tilde S(\bk)$
is the Fourier transform of the density-density correlation.  Close to
a reciprocal lattice vector,
\begin{equation*}
k_z \equiv q_z, \quad 
\bk_\perp \equiv \bG+\bq_\perp, \quad \bq_\perp \ll G,
\end{equation*}
the structure factor becomes
\begin{eqnarray}
\tilde S(\bG+\bq_\perp,q_z) &=& \int_z \sum_\bX e^{iq_zz+i\bq_\perp\bX}
\overline{\langle e^{i(\bG+\bq_\perp)(\bu(\bX,z)-\bu(\bN,0))}\rangle}\nn\\
&\simeq& \int d^3r  e^{iq_zz+i\bq_\perp\bx} C_{\bG}(\br).
\end{eqnarray}
The scattered intensity is thus described by the Fourier transform of
the translational order correlation $C_{\bG}(\br)$ defined in
Eq. (\ref{Scorr}).  The Fourier transformation can be done numerically
for general $\kappa$ but one can see easily that the power law decay
of $C_{\bG}(\br)$ is slow enough to effect powerlaw divergence of
$\tilde S(\bG+\bq_\perp,q_z)$ for small $\bq_\perp$. For the limiting
cases $c_{11}=c_{66},\kappa=1$ and $c_{11}\gg c_{66},
\kappa\backsimeq 0$, of which the latter is realized physically close
to $H_{c_2}$, the structure factor reads
\begin{equation}
\label{structurefactor-limits}
\tilde S(\bG+\bq_\perp,q_z)\propto \left(\bq_\perp^2 + {{c_{44}}
\over{c_{66}}}q_z^2\right)^{(-3+\eta_\bG(\kappa))/2}.
\end{equation}
As a consequence of the quasi-long-range order in the asymptotic regime
beyond the scale $L_a$, the structure factor diverges for small
($\bq_\perp,q_z$) leading to Bragg peaks.  Giamarchi and Le Doussal
thereafter named the weakly disordered flux line lattice a {\em Bragg-glass}.

In experiments, Bragg peaks and their disappearence for higher
magnetic fields due to a melting of the flux line lattice have in fact
been observed in BSCCO \cite{Cubitt+93,Zeldov+95}.  However,
quantitative details of the divergence cannot be resolved putting our
prediction of nonuniversality of $\eta_\bG$ beyond the precision of
today's experimental diffraction techniques. But there exist recent
experimental results which are based on a more microscopic way to
determine the structure of flux line lattices. Kim {\em et al.} use a
scanning electron microscope to obtain a spatial map of the flux line
displacements itself over a region containing about $5\cdot 10^5$
lines \cite{Kim+99}. Due to the small coherence length of $\approx
20\AA$ in BSCCO we have $\xi\ll a$ so that only the RM and A regime
can be observed in principle. Experimental results for the mean
squared relative displacement of lines are available in the RM regime,
whereas the asymptotic regime is beyond the experimental limit. Kim
{\em et al.} measure in this regime a roughness exponent of
$\zeta_{RM}=0.22$. This result is in reasonable agreement with our
first order $\epsilon$ expansion result of $\zeta_{RM}\approx 0.18$ in
3D.

%---------------------------
%---------------------------
\section{Discussion and conclusion}
\label{discussion_and_conclusion}
%---------------------------
%---------------------------

For the first time, the three dimensional Abrikosov lattice in the
presence of point disorder has been treated within a renormalization
group procedure including all of the elastic modes. Triangular
symmetry complicates the calculation of large scale 3D correlation
functions in their anisotropic dependency on position arguments. These
technical difficulties could be overcome and we can write correlations
in their full anisotropy.  More important however, translational
quasi-long-range order in impure type-II superconductors is described
by a -- contrary to previous claims -- {\em nonuniversal} power-like
decay of the order parameter correlations. In particular, the
decay-exponent $\eta_{\bf G}$ depends (if weakly) on the ratio
$\kappa= c_{66}/{c}_{11}$ of the elastic constants.

In isotropic superconductors at low temperatures, where flux lines
interact via central forces, one has $0\le\kappa\le 1/3$. $\kappa\sim
1/3$ for $\lambda\le a$, i.e. for fields close to $H_{c_1}$, and
$\kappa \rightarrow 0$ for $H \rightarrow H_{c_2}$.  For most of the
field region $\kappa \approx \phi_0/16\pi\lambda^2B$,.  Thus, an
increase of the external field from $H_{c_1}$ to $H_{c_2}$ should
result in an increase of $\eta_{{\bf G}}$ and a decrease of
$\zeta_{RM}$.  Numerically, the effect is small, since $\eta_{{\bf
    G}_0}$ ranges from $1.145$ to $1.159$ and $\zeta_{rm}$ from
$0.1745$ to $0.1763$ in this $\kappa$-range. Thus it will probably be
hard to detect this effect.  However, we are likely to have suppressed
some of the nonuniversality in our calculation when we extended the
$z$-direction to 2 dimensions.  In the propagator the effect of
variations in $c_{11}$ and $c_{66}$ is thus reduced by the more
heavily weighted $c_{44}\bq_z^2$-term. The behaviour of 2D disordered
triangular lattices supports this point of view.  Here, Carpentier and
Le Doussal obtain nonuniversal large scale behaviour that is much
more pronounced than in our higher dimensional case
\cite{Carpentier+97}.  The small quantity of the calculated effect
should thus not mislead to underestimate its qualitative relevance.

Different from usual RG approaches we observed the full flow of
renormalization of the the interaction.  We thus gained in addition to
the fixed point dominated large scale behaviour information on the
intermediate length scales. We find a crossover of the structural
correlation functions from a Larkin-regime, where perturbation theory
applies, to the random manifold regime and eventually to the
asymptotic Bragg glass regime.  Out of one Hamiltonian this confirms
most clearly the physical picture that had originated over many years
from different approximation strategies to this prominent physical
system. For the random manifold regime, where fluxlines explore many
minima in the energy landscape but do not yet compete against each
other, we could extract the roughness exponent numerically from the
flow. This is valuable in itself as it can be compared only to an
estimated exponent from scaling arguments.  Moreover it also shows
dependency on elastic constants, i.e., is nonuniversal.\\ Dislocations
are excluded in our description. Whereas the stability of the
Bragg-glass phase against these topological defects had not always
been commonly agreed upon, today the position of the Bragg-glass in
the phase diagram is well established.  In the $H-T$ plane at not too
large fields, weak disorder -- the definition of which can be cast
into a Lindemann criterion -- has the Abrikosov phase of the pure
phase diagram become the Bragg-glass.  It is bounded by a first order
transition melting line with negative slope, just like the Abrikosov
phase in the pure system. This line ends at a critical point.  For
larger fields and small temperatures the Bragg-glass is bounded by the
proliferation of vortices that destroy quasi-long-range order
\cite{Gingras+96,Kierfeld+97,Ertas+96,Carpentier+96,FisherDS97,Kierfeld98}.
The exact nature of the large field state -- it may be called a
dislocated vortex glass -- is not clearly understood, nor is the
transition (or mere crossing) into it from the Bragg-glass. Berker set
up a criterion for the shift of a first order transition to a second
order one by disorder \cite{Berker93}.  For a sharp domain boundary at
the coexistence point he finds a pure first order transition to be
stable against disorder fluctuations above 2D. This is applicable to
the melting line where a jump in the FL density, which plays the role
of a scalar order parameter, assures the sharp boundary and a first
order transition is observed both in pure and impure samples
\cite{Soibel+00}.

%%%%%%%%%%%%%%%%%%%%%%%%%%%%

We would like to thank very much H. E. Brandt, M. Kardar and S.
Scheidl for valuable discussions.  Financial support was obtained
through Deutsche Forschungsgemeinschaft under grant No. EM70/1-3
(T.E.).

\newpage

\begin{appendix} 
%---------------------------
\setcounter{equation}{0}
%---------------------------
%---------------------------
%---------------------------
\section{Replica pinning energy}
\label{Replica_pinning_energy}
%---------------------------
%---------------------------
We want to write the replica pinning energy in compact form starting
from the pinning energy expression (\ref{h_dis2}). The
$\bG\ne\bN$-terms are the relevant ones, as we have argued.
\begin{eqnarray}
\label{replica_appendix1}
\lefteqn{
\cH_{dis}^n[\{\bu^a\}]=}\hspace{0cm}
\nn\\&=&
-{1\over2T}\overline{\Big[\rho_0\int_\br V_{dis}(\br)\sum_{\bG\ne\bN}\sum_{a=1}^n e^{i\bG[\bx-\bu^a(\br)]}\Big]^2}
\nn \\
&=&
-{1\over2T}\rho_0^2\sum_{a,b}\int_{\br\br'}\Delta_{\xi}(\bx-\bx')\delta(z-z')
\nn\\
&&\hspace*{1cm}\times
\sum_{\bG\bG'}e^{i\bG[\bx-\bu^a(\br)]+i\bG'[\bx'-\bu^b(\br')]}.
\end{eqnarray}
We focus on the $\bx,\bx'$-dependency and suppress the $z$-index of
the displacement fields for shorter notation.

\begin{eqnarray}
\lefteqn{
\sum_{\bG\bG'}\int_{\bx\bx'} \!\!\!\Delta_{\xi}(\bx-\bx')e^{i\bG[\bx-\bu^a(\bx)]+i\bG'[\bx'-\bu^b(\bx')]}=}
\nn\\
&=&
\sum_{\bG}\int_{\bx} e^{-i\bG[\bu^a(\bx)-\bu^b(\bx)]}
\sum_{\bG'}e^{i[\bG+\bG'][\bx-\bu^b(\bx)]}
\nn\\
&&
\times \int_{\bar\bx}\Delta_{\xi}(\bar\bx)e^{i\bG'[\bar\bx-(\bu^b(\bx+\bar\bx)-\bu^b(\bx))]}
\nn\\
&\backsimeq&
\sum_{\bG}\int_{\bx} e^{-i\bG[\bu^a(\bx)-\bu^b(\bx)]}
\nn\\
&&
\times\Big(\tilde\Delta(\bG)+\sum_{\bG'\ne-\bG}e^{i[\bG+\bG'][\bx-\bu^b(\bx)]}\tilde\Delta(\bG')\Big)
\nn\\
&\backsimeq&
\int_{\bx}\sum_{\bG} \tilde\Delta(\bG) e^{-i\bG[\bu^a(\bx)-\bu^b(\bx)]}.
\end{eqnarray}
$\tilde{\Delta}(\bk)$ is the Fourier transform of $\Delta_\xi(\bx)$,
it is nonzero over a region of size $\xi^{-1}$.  For the first
approximation the slow variation of $[\bu^b(\bx+\bar\bx)-\bu^b(\bx)]$
with $\bar\bx$ is used. In the region $|\bar\bx|\le\xi$, which is the
one contributing to the integral, its variation is negligable versus
$\bar\bx$. In the last step, rapidly oscillating terms are neglected
versus the constant $\tilde{\Delta}(\bG)$ as they are both multiplied
by the slowly oscillating $e^{-i\bG[\bu^a(\bx)-\bu^b(\bx)]}$ and
integrated over.\\ This gives the replica pinning energy as stated in
equation (\ref{repham-complete})
\begin{eqnarray}
\cH_{dis}^n[\{\bu^a\}]&=&-{1\over2T}\sum_{a,b}\int_{\br}\rho_0^2\sum_{\bG\ne\bN} \tilde\Delta(\bG) e^{-i\bG[\bu^a(\br)-\bu^b(\br)]}
\nn \\
&\equiv&
-{1\over2T}\sum_{a,b}\int_{\br}R(\bu^a(\br)-\bu^b(\br)).
\end{eqnarray}
In this version of the replica pinning energy, the effective disorder
correlator $R(\bu)$ is invariant under the full symmetry of the
triangular lattice.  This is because pinning energy depends only on
the positions $\bX+\bu_\bX(z)$ of the vortices.  Redistributing the
fluxlines to reference positions, i.e., relabelling, may thus not
show.  $\cH_{dis}$ in Eq. (\ref{Hdis}) is consistently invariant under
$\bu_\bX(z)\rightarrow\bu_{\bX+\bY}(z)+\bY$ with $\bY$ a constant
lattice vector. This reads $\btu(\btx)\rightarrow\btu(\btx)+\bY$ after
substitution (\ref{substitute}) and explains the discrete
translational invariance of the Fourier sum $R(\bu)$.\\ The point
group symmetries also arise from the invariance of disorder energy
under a change of the `original' positions of the fluxlines in the
pure system.  Let us again resume the distinct notation for the field
before ($\bu(\bx)$) and after ($\btu(\btx)$) substitution
(\ref{substitute}).
\begin{equation}
\label{sym1}
\btu(\btx)\rightarrow \bD_{60^\circ}\btu(\btx) 
\end{equation}
is not an exact symmetry of the pinning energy, it is rather the
transformation
\begin{equation}
\label{sym2}
\btu(\btx)\rightarrow \bD_{60^\circ}\btu(\btx)+(1-\bD_{60^\circ})\btx,
\end{equation}
i.e.,
\begin{equation}
\btx-\btu(\btx)\rightarrow \bD_{60^\circ}(\btx-\btu(\btx))\nn
\end{equation}
that leaves $\cH_{dis}$ unchanged, as can be seen in
(\ref{with_poisson}) (the rotation is absorbed in the reciprocal
lattice, that is mapped onto itself as $\bD_{60^\circ}$ is a lattice
symmetry).  For the local term $\bG=-\bG'$ that we keep for the final
compact form of the pinning energy, symmetry (\ref{sym2}) becomes
symmetry (\ref{sym1}). It reads for $\{\bx,\bu(\bx)\}$
\begin{equation}
\label{sym3}
\bu(\bx)\rightarrow \bu(\bD_{60^\circ}^{-1}\bx)+(\bD_{60^\circ}^{-1}-1)\bx.
\end{equation}
The disorder energy in Eq. (\ref{Hdis}) is transformed as
\begin{eqnarray*}
\label{sym4}
\sum_{\bX} \int\! \d z^{d-2} \!\:V_{dis}\left(\bX+\bu\left(\bX \right),z\right)
\rightarrow&&\\
\sum_{\bX} \int\! \d z^{d-2} \!\:V_{dis}\left(\bD_{60^\circ}^{-1}\bX+\bu\left(\bD_{60^\circ}^{-1}\bX\right),z\right)&&
\end{eqnarray*}
and is invariant as the lattice can be relabelled $\bX\rightarrow
\bD_{60^\circ}\bX$.  Our starting Hamiltonian (\ref{repham-complete})
for the RG is written in the quantities with the tilde.  Correlations
calculated above are thus also expressed in $\{\btx,\btu\}$. For any
roughness with exponent smaller than one however, correlations of
$\btu$ coincide with the ones of $\bu$. This justifies our dropping of
the tilde in most of the treatment above.

%---------------------------
%---------------------------
\section{Integrals for the equation of flow}
%---------------------------
%---------------------------
\label{appendix-eof}
In this appendix the coefficients $\bM$ of equation (\ref{eof_raw}) are calculated.
\begin{eqnarray*}
\bM^{\alpha \beta,\gamma \delta}&=&\d l^{-1}
           \int_\bq^> \cG_{\alpha \beta}(\bq) \cG_{\gamma \delta}(-\bq)
\end{eqnarray*}
with
\begin{eqnarray*}
\cG_{\alpha \beta}(\bq) \cG_{\gamma \delta}(-\bq)&=&
\bP_{\alpha \beta}^L \bP_{\gamma \delta}^L 
  \left(\cG^L\right)^2 +
        \bP_{\alpha \beta}^T \bP_{\gamma \delta}^T 
          \left(\cG^T\right)^2\\
 && +   \left( \bP_{\alpha \beta}^L \bP_{\gamma \delta}^T +
        \bP_{\alpha \beta}^T \bP_{\gamma \delta}^L \right).
                \cG^L \cG^T.
\end{eqnarray*}
The projectors $\bP^{L,T}$ are given after equation (\ref{hamq}). 
Take $(\alpha \beta),\: (\gamma \delta)$ as column- and row-indices of a
$4 \times 4$ array. The indices shall run through $(xx, xy, yx, yy)$.
Then
\begin{eqnarray*}
\int_\bq^>\bP_{\alpha \beta}^L \bP_{\gamma \delta}^L 
  \left(\cG^L\right)^2 &=&
     {1\over 8}
        \left( \begin{array}{cccc}
        3&0&0&1\\
        0&1&1&0\\
        0&1&1&0\\
        1&0&0&3 \end{array}\right)\int_\bq^> \cG_L^2. 
\end{eqnarray*}
Here the relations
\begin{eqnarray*}
&&\int_\bq^> \cG_L^2 {q_y^4\over q_\perp^4}=
  3 \int_\bq^> \cG_L^2 {q_x^2 q_y^2\over q_\perp^4},\\
&2&\int_\bq^>\left[ \cG_L^2 {q_y^4\over q_\perp^4}+
     \cG_L^2 {q_x^2 q_y^2\over q_\perp^4}\right]=\\
&=&\int_\bq^> \cG_L^2{1\over{q_\perp^4}}(q_y^4+2q_y^2q_x^2+q_x^4)
=\int_\bq^> \cG_L^2
\end{eqnarray*}
have been used. Similarily, one obtains
\begin{eqnarray*}
\int_\bq^> \bP_{\alpha \beta}^T \bP_{\gamma \delta}^T 
          \left(\cG^T\right)^2 &=&
         {1\over 8}
        \left( \begin{array}{cccc}
        3&0&0&1\\
        0&1&1&0\\
        0&1&1&0\\
        1&0&0&3 \end{array}\right)\int_\bq^> \cG_T^2
\end{eqnarray*}
and
\begin{eqnarray*}
\int_\bq^>\bP_{\alpha \beta}^T \bP_{\gamma \delta}^L 
                \cG^L \cG^T
&=&{1\over 8}
        \left( \begin{array}{cccc}
        2&0&0&6\\
        0&-2&-2&0\\
        0&-2&-2&0\\
        6&0&0&2 \end{array}\right)\int_\bq^> \cG_T \cG_L.
\end{eqnarray*}
This gives $\bM^{\alpha \beta,\gamma \delta}$ as stated
in (\ref{M-computed})
\[
\bM={1 \over 8} 
        \left( \begin{array}{cccc}
        3 I_1 + 2 I_2 & 0 & 0 &  I_1 + 6 I_2 \\
        0 &  I_1 - 2 I_2 & I_1 - 2 I_2& 0 \\
        0 & I_1 - 2 I_2& I_1 - 2 I_2& 0 \\
        I_1 + 6 I_2 & 0 & 0 & 3 I_1 + 2 I_2
        \end{array} \right)
\]
with
\[
\begin{array}{cc}
I_1 \equiv
      dl^{-1}  \int_\bq^> \left( \cG^2_T + \cG^2_L \right) &\quad
I_2\equiv
        dl^{-1} \int_\bq^> \cG_T \cG_L 
\end{array}.
\]
These integrals are evaluated exactly in $d=4$ to
\[
I_1= {1\over {8\pi^2}}
        {{1+\kappa}\over {c_{44} c_{66}}}\quad
I_2=\frac{1}{8\pi^2 c_{44}c_{11}}
        \frac{\ln \kappa}{\kappa-1}
\]
with $\kappa={{c_{66}}/{c_{11}}}$.

%---------------------------
%---------------------------
\section{Fourier transform of the anisotropic propagator}
\label{fourier_transform}
%---------------------------
In this appendix the displacement correlation functions shall be
calculated from the propagator $\overline{\langle u_\alpha(\bq)
  u_\beta(\bq') \rangle}$ in Eq. (\ref{propagator}). The Fourier
transformation reads
\[
B_{\alpha \beta}(\br)= 2 \int_{\bq\bq'}^{L_a^{-1}} \!\!\overline{\langle 
u_\alpha(\bq)u_\beta(\bq')\rangle} (1-\cos \bq\br)
.\]
We integrate in $d=4$ with the $z$-direction extended to a 2D subspace. The 
$xx$-displacement correlations then are
\begin{eqnarray*}
\lefteqn{
B_{xx}(\bx, \bz)=
}\hspace{0cm} \\
&=&C
\int \!\d^4\!q\: \left(1-\cos(\bq_z \bz + \bq_\perp \bx)\right)
\\&&\times
\left[\frac{q_x^2}{q_\perp^2}\cG_L^2 + \frac{q_y^2}{q_\perp^2}\cG_T^2\right]
\\
&=&C\int\! \d^4\!q\: \left\{1-\cos(\bq_z \bz)\cos(\bq_\perp \bx)
+  \sin(\bq_z \bz)\sin(\bq_\perp \bx)\right\}\\
&&\times
\left[\frac{q_x^2}{q_\perp^2}\cG_L^2 + \frac{q_y^2}{q_\perp^2}\cG_T^2\right]
\nn\\
&=&C\int\! \d^2\!q_z\left\{\int\! \d q_\perp q_\perp
\cG_L^2 \int_0^{2\pi}\! \d\phi \left[1-\cos(\bq_z \bz)\right.\right.
\\&&\times
  \cos(q_\perp x\cos\phi+q_\perp y\sin\phi)
\\&&+\left.
  \sin(\bq_z \bz)\sin(q_\perp x\cos\phi+q_\perp y\sin\phi)\right]\cos^2\phi 
\\&&\left.
 + \cG_T\mbox{-analogue}\right\}
\end{eqnarray*}
with $C=2 \Delta^* (2\pi)^{-4}$.
The $\sin(\bq_z \bz)$-term does not
contribute as it is antisymmetric in $\bq_z$. So we have
\begin{eqnarray*}
\lefteqn{
B_{xx}(\bx, \bz)=
}\hspace{0cm} \\
&=&
C\pi\int\! \d^2q_z \int\! \d q_\perp q_\perp \Big\{\cG_L^2\Big[
1-\cos(\bq_z \bz)
\\&& \times 
\Big(J_0(q_\perp |\bx|)+
  \frac{y^2-x^2}{\bx^2}J_2(q_\perp |\bx|)\Big)\Big]
\\&&+\cG_T^2\Big[
1-\cos(\bq_z \bz)
\\&&\times
\Big(J_0(q_\perp |\bx|)+
  \frac{x^2-y^2}{\bx^2}J_2(q_\perp |\bx|)\Big)\Big]\Big\}
\end{eqnarray*}
with $J_n$ the Bessel function of the $n$-th kind. We used 
\begin{eqnarray*}
\lefteqn{
\int_0^{2\pi}\!\d\phi \cos(q_\perp x\cos\phi+q_\perp y\sin\phi)\cos^2\phi
}\hspace{0cm}\\
&=& J_0(q_\perp |\bx|)+\pi \frac{y^2-x^2}{\bx^2}
   J_2(q_\perp |\bx|)
\end{eqnarray*}
and its analogue for the $\cG_T$-term.
With $\bq_z$ written in spherical coordinates as well, the
$q_z$-integration is done easily using standard tables,
\begin{eqnarray*}
\lefteqn{
B_{xx}(\bx, \bz)=
}\hspace{0cm} \\
&&
2\pi^2 C \int\! \d q_\perp q_\perp \int\! \d q_z q_z \left\{\cG_L^2(q_\perp,q_z)
\left[1-J_0(q_z z) \right. \right.
\\&& \times \left. \left.
\left(J_0(q_\perp |\bx|)+
  \frac{y^2-x^2}{\bx^2}J_2(q_\perp |\bx|)\right)\right]\right.\nn\\
&&+\: \cG_T^2(q_\perp,q_z)
\left[1-J_0(q_z z) \right.
\\&& \times \left. \left.
\left(J_0(q_\perp |\bx|)+
  \frac{x^2-y^2}{\bx^2}J_2(q_\perp |\bx|)\right)\right]\right\}\\
&=&\frac{\pi^2 C}{c_{44}} \int\! \d q_\perp \left\{\frac{1}{c_{11}}\left[
\frac{1}{q_\perp}-z_l K_1(z_l q_\perp) \right.\right.\\
&&\times \left. \left.
\left(J_0(q_\perp |\bx|)+
  \frac{y^2-x^2}{\bx^2}J_2(q_\perp |\bx|)\right)\right]\right.+
\end{eqnarray*}
\begin{eqnarray*}
&&+
\frac{1}{c_{66}}\left[
\frac{1}{q_\perp}-z_t K_1(z_t q_\perp) \right. \\
&&\times \left. \left. \left.(J_0(q_\perp |\bx|)+
  \frac{x^2-y^2}{\bx^2}J_2(q_\perp |\bx|)\right)\right]\right\}.
\end{eqnarray*}
$K_n$ is the modified 
Bessel function of the $n$-th kind.
We have once more rescaled the coordinates according to $z_l=z \sqrt{c_{11}/
c_{44}}$, $z_t=z \sqrt{{c_{66}}/{c_{44}}}$.
The third terms in both the longitudinal and transverse part is not 
pathological ($K_1(q)\sim q^{-1},\: J_2(q)\sim q^2 \: \mbox{ for } q 
\rightarrow 0$) and can easily be integrated. Then 
\begin{eqnarray}
\label{xy-last-but-one}
\lefteqn{
B_{xx}(\bx, \bz)=
}\hspace{0cm} \nn\\
&&\frac{\pi^2 C}{c_{44}c_{11}}\Bigg\{
\int\! dq_\perp\left[
{1\over {q_\perp}}-z_lK_1(z_lq_\perp)J_0(q_\perp|\bx|)\right]\nn \\
&&+\left.
\frac{x^2-y^2}{2\bx^2}\left(1-\left({|\bx|\over z_l}\right)^{-2}
\ln\left[1+\left({|\bx|\over z_l}\right)^2\right)\right]\right\}\nn\\
&&+
\frac{\pi^2 C}{c_{44}c_{66}}\left\{\int\! \d q_\perp\left[
{1\over {q_\perp}}-z_tK_1(z_lq_\perp)J_0(q_\perp|\bx|)\right]\right.\nn\\
&&+\left.
\frac{y^2-x^2}{2\bx^2}\left(1-\left({z_t^2\over |\bx|^2}\right)
\ln\left[1+{|\bx|^2\over z_t^2}\right]\right)\right\}.
\end{eqnarray}
To separate
the cancelling divergences from
non-diverging `geometric terms' we introduce the function
\[
\label{f-integral}
f(\gamma)=
\int_0^{|\bx|L_a^{-1}}\!\d q\left\{{1\over {q}}-\gamma^2K_1(\gamma^2q)J_0(q)\right\}
\]
which yields the integral in Eq. (\ref{xy-last-but-one}) for $\gamma^2=z/|\bx|$.
This function can be written as 
\[
f(\gamma)=f(0)+\int_0^\gamma f'(t)
\]
with
\begin{eqnarray*}
\label{f-gamma}
&&f(0)=\int_0^{|\bx|L_a^{-1}}\!\d q {1\over q}(1-J_0(q))=\ln {|\bx|\over {L_a}}+\cO(1)
\\
&&f'(t)= 2t^3\int_0^{|\bx|L_a^{-1}} \!\!\!\!\!\d q\, q K_0(t^2q)J_0(q)\qquad 
\\
&&= {2t^3} 
\frac{1+{|\bx|\over {L_a}}J_1({|\bx|\over {L_a}})K_0(t^2{|\bx|\over {L_a}})
-t^2{|\bx|\over {L_a}}J_0({|\bx|\over {L_a}})K_1(t^2{|\bx|\over {L_a}})}
{1+t^4}.
\end{eqnarray*}
Integrating the first term of $f'(t)$ gives
\[
\int_0^\gamma \!\d t\frac{2t^3}{1+t^4}= {1\over 2}\ln (1+\gamma^4)
,\]
whereas the latter two yield vanishing integrals for large $|\bx|$.
We now have 
\begin{eqnarray*}
f(\gamma=\sqrt{z/|\bx|})&=&\ln {|\bx|\over{L_a}}  +{1\over 2}\ln \left[1+\left({z\over{|\bx|}}\right)^2\right]+ \cO(1)\\
&=&{1\over 2}\ln \left[\left({|\bx|\over{L_a}}\right)^2+
  \left({z\over{L_a}}\right)^2\right]+ \cO(1).
\end{eqnarray*}
With the definitions $\kappa={c_{66}}/{c_{11}}$,
$h(t)=t^{-2}\ln(1+t^2)$ and considering the rescaling for the equation of flow
(\ref{Rrescaling}), (\ref{Is}), as to write in terms of the
the dimensionless $\tilde{\Delta}^*$ from the numerical solution of
the RG equations, we finally have
\begin{eqnarray*}
\lefteqn{
\overline{\langle [u_x(\bx,\bz)-u_x(\bN,\bN)]^2\rangle}=}\hspace{.5cm}\\
&&\frac{\tilde{\Delta}^*a^2}{1+\kappa}\left\{
\ln \left[\left({|\bx|\over{L_a}}\right)^2+
  \left({z_t\over{L_a}}\right)^2\right]\right.\\ 
&&\left.+\kappa\ln \left[\left({|\bx|\over{L_a}}\right)^2+
  \left({z_l\over{L_a}}\right)^2\right]\right.\\ 
&& + \left.\frac{y^2-x^2}{2\bx^2}\left(1-h\left({|\bx|\over z_l}\right)
-\kappa\left[1-h\left({|\bx|\over z_t}\right)\right]\right)\right\}.
\end{eqnarray*}
All nonconverging terms for $z,|\bx| \rightarrow \infty$ have been
retained. The $yy$-corre\-la\-tions are obtained by $\kappa
\rightarrow \kappa^{-1}$, $(x,y)\rightarrow (y,-x)$ as explained in
Section \ref{symmetries}. The mixed $xy$-corre\-la\-tions are
calculated similarily and read
\begin{eqnarray*}
\lefteqn{
B_{xy}(\bx, \bz)=}\\
&=&\overline{\langle [u_x(\br)-u_x(\bN)][u_y(\br)-u_y(\bN)]
\rangle}\\
&=&\pi^2 C\frac{xy}{\bx^2}\frac{1}{c_{44}c_{66}}\left\{
h\left(\frac{|\bx|}{z_t}\right)-1-\kappa\left[h\left(\frac{|\bx|}{z_t}\right)
-1\right]\right\}\\
&=&
2\frac{\tilde{\Delta}^*a^2}{1+\kappa} 
\frac{xy}{\bx^2}\left\{
h\left(\frac{|\bx|}{z_t}\right)-1-\kappa\left[h\left(\frac{|\bx|}{z_t}\right)
-1\right]\right\}.
\end{eqnarray*}

\end{appendix}

%%%%%%%%%%%%%%%%%%%%%%%%%%%%
% references %%%%%%%%%%%%%%%

\end{document}